\documentclass[apjpt4]{aastex}

\shorttitle{Flattened Envelope around L1157}
\shortauthors{Chiang et al.}

\begin{document}

\title{Probing the Protostellar Envelope around L1157: the Dust and Gas Connection} 
\author{Hsin-Fang Chiang\altaffilmark{1}, 
Leslie W. Looney\altaffilmark{1},
John J. Tobin\altaffilmark{2}, Lee Hartmann \altaffilmark{2} 
}

\altaffiltext{1}{Department of Astronomy,
University of Illinois at Urbana-Champaign, 
1002 West Green Street, Urbana, IL 61801;
hchiang2@uiuc.edu}
\altaffiltext{2}{Department of Astronomy,
University of Michigan, Ann Arbor, MI 48109}  

\begin{abstract}

We present observations of the Class 0 protostar L1157-mm using the
Combined Array for Research in Millimeter-wave Astronomy (CARMA)
in 3 mm dust continuum and N$_2$H$^+$ line emission.
In the N$_2$H$^+$ line, we detect a large-scale envelope
extended over a linear size of $\sim$20,000AU flattened in the direction 
perpendicular to the outflow.  This N$_2$H$^+$ feature coincides with the
outer envelope seen in the 8 $\mu$m extinction by Looney et al. 
%8 $\mu$m extinction feature seen in Looney et al. (2007), showing the
%outer envelope that is too dim to be detected in dust continuum.
Meanwhile, the dust continuum traces the compact, nearly spherical
structure of the inner envelope, where N$_2$H$^+$ becomes depleted.
This highly flattened N$_2$H$^+$ envelope also shows dynamical signatures 
consistent with gravitational infall in the inner region, but 
a slow, solid-body rotation at large scales.     
%; though the structure is not rotationally supported.
This flattened structure is not a rotationally supported
circumstellar disk; instead, it resembles a prestellar core both
morphologically and kinematically, 
representing the early phase of a Class 0 system.  
%that properties of the preceding prestellar core can be seen at large scales.  
%, suggesting that it can be the remnant
%from the dynamical processes of early core formation.
In this paper, we construct a simple model to interpret both the dust
continuum and N$_2$H$^+$ emission and suggest a possible dynamical
scenario for the overall properties of the envelope.

%observations of N2H+(column density, morphology, infrared obs...); dynamics study from gas; comparing dust and gas

%By comparing the N$_2$H$^+$ feature with the 8 $\mu$m extinction, 
%we examine the N$_2$H$^+$ abundance along the flattened envelope. 
%%; enhancement of the N$_2$H$^+$ abundance near the central protostar is suggested.  

\end{abstract}


\maketitle

\section{Introduction}

Significant progress has been made both theoretically and observationally
toward understanding low-mass star formation in the last few decades,
and an evolutionary sequence has been posited
%\citep[e.g., see reviews of][]{Andre2000,McKee2007}. 
\citep[e.g.,][]{Andre2000,McKee2007,Evans2009}. 
The youngest protostars, the so-called  Class 0 sources, form from
the preceding stage of a prestellar core and are deeply embedded in
their natal collapsing envelope.
%Gravitational collapse happens in the prestellar cores, 
%where protostars form deeply inside their natal collapsing envelopes. 
% and undergoing gravitational collapse. 
%with well defined bipolar outflows. 
As a link from prestellar cores to young protostars, 
%, Class 0 objects, especially their envelopes, 
these envelopes contain valuable physical clues to the
initial conditions of the collapse process. 
%, which provide better insight of protostellar evolution.   
%Significant progress has been made towards understanding low mass star
%formation in the last few decades both theoretically and observationally.
%For example, a sketch of low mass star formation evolution has been
%suggested. The possibly transient starless cores and the subgroup of
%self-gravitating prestellar cores, objects without detected infrared
%sources, are the condensations in dense clouds where protostars are born.
%Dynamical processes, such as gravity, rotation, magnetic fields, and
%turbulence, in combination likely induce (wrong?) the infall motion
%of collapse in these condensations that lead to the protostars (ref?).
%The so-called Class 0 protostars are the youngest protostars and deeply
%embedded within their natal envelopes (ref?). Most material in the
%envelope collapses onto the protostellar disks with a bipolar region
%evacuated by the widening outflows.  The protostar continues to evolve
%through the Class I, II, and III stages with a possibly planet-forming
%circumstellar disk. As a link from prestellar cores to young protostars,
%Class 0 objects and especially their envelopes contain valuable physical
%clues of the collapse process to better understand the overall evolution.
While single-dish telescope observations alone are usually limited by  
large beam size, interferometry has allowed high resolution studies of the 
Class 0 envelope.  In particular, interferometric observations 
combined with dust modeling using theoretically predicted structure can 
probe the physical properties of the inner envelope 
\citep[e.g.,][]{Looney2003,Chiang2008}.
Unfortunately, dust emission from the outer 
envelope is usually difficult to detect due to interferometer 
insensitivity to the large-scale structure or the low-surface brightness of 
the extended emission. Molecular lines, on the other hand, trace specific 
components of the envelope and can detect regions of lower density. 
Further, as the line emission is dispersed over many velocity channels, 
the structures are less extended than the equivalent dust emission so 
less affected by interferometer resolving-out issues.  
%, and are somewhat less affected by interferometer resolving-out issues 
%as the structure is dispersed over many velocities.  
However, uncertainty in abundances and chemistry in the envelope can make
molecular line interpretation difficult.

Since the first detection in \citet{Turner1974} and confirmation in 
\citet{Thaddeus1975}, N$_2$H$^+$ has been a well-known interstellar molecule
%, especially for studying quiescent cores and young stellar objects.
and an excellent tracer of dense cores 
\citep[e.g.,][]{Caselli2002,Chen2007}.  % Fontani2006 
With a critical density of around 10$^5$ cm$^{-3}$ \citep[e.g.,][]{Daniel2006}, 
N$_2$H$^+$ is sensitive to the outer envelope around Class 0 
young stellar objects (YSOs). 
Also, 
N$_2$H$^+$ has a low depletion rate and remains in the gas phase when other
molecules such as CO have depleted onto grains \citep{Bergin1997}.
The reason for this was thought to be the low binding energy of its parent
molecule N$_2$, but recent experiments have shown that the binding energy
of N$_2$ is not much lower than that of CO \citep{Oberg2005}.  
In fact, most of the nitrogen is in the atomic form, which 
has a binding energy lower than the molecular form, 
and therefore making N$_2$H$^+$ harder 
to deplete  \citep{Maret2006}.
%instead of molecular form 
%the atomic nitrogen has a low binding energy 
%Nonetheless, N$_2$H$^+$ is an excellent tracer of dense cores 
%%usually exists as the gaseous counterpart of the dust continuum
%and commonly found in cores and protostars   
%\citep[e.g.,][]{Caselli2002,Chen2007}.  % Fontani2006 
In this paper, we use N$_2$H$^+$ as an envelope tracer to reveal 
the gas in the outer protostellar envelope, as well as pursue a comparison 
between gas and dust.  
%---------------------------------------------

%Actually, ion does not deplete onto grains, it's their parent molecules that 
%freezeout/condense/deplete onto grain surface, indirectly reducing the
%abundance of the gas molecules at high density region.

%---L1157
While most Class 0 protostellar envelopes have rather complex structures, 
L1157-mm (also known as L1157-IRS or IRAS 20386+6751) 
in the Cepheus flare \citep[e.g.,][]{Kun2008,Kirk2009}
%is a nice example of isolated star formation, with 
has a highly flattened and relatively symmetric envelope   
%In \citet{Looney2007}, a flattened structure 
2\arcmin~across seen by 8 $\mu$m absorption  
%like a textbook example 
%show no hint of multiple system. 
\citep{Looney2007,Tobin2009}. 
%observations have shown the variety of envelope morphology (Tobin2009). 
%Simple assumptions of symmetry may not be applicable to most of the objects. 
%No clear evidence of binary is found in this system.  
The protostar is in the Class 0 stage with an estimated age of 8-150 kyr  
\citep[][based on bolometric temperature, bolometric luminosity, 
and submillimeter luminosity]{Froebrich2005}.  
Different components of the dust emission, 
including a central compact core and an extended envelope, 
have been observed \citep[e.g.,][]{Gueth1997,Gueth2003,Beltran2004}.  
Also, large-scale outflows have been detected by various molecules, 
suggesting an inclination angle of 80\degr~\citep{Gueth1996} and 
% meaning that the system is nearly edge-on \citep{Gueth1996}.  
a kinematic age of 15 kyr  derived from the oldest pair of outflow clumps
\citep[][and reference therein]{Bachiller2001,Arce2008}.
%(Bachiller1993,1995,1997,2001)(Arce2008).
%kinematic age: (Bachiller2001) 
Furthermore, detection of methanol has suggested active accretion in the 
embedded circumstellar disk \citep{Goldsmith1999L1157,Velusamy2002}.
%may be a detection of the pseudodisk \citep{Galli1993}. 
%Known to be a complex system of infall, outflow, and rotation, the 
%kinematics of the system can only be probed using molecular lines.  
%All above suggest that L1157 is a complex system with  
With features of a typical Class 0 young stellar object,  
L1157 seems to be a good example of isolated low-mass star formation. 

%kinematic info is needed to understand the underlying physical 
%properties of the flattened feature. 
%Kinematically, infall + outflow + rotation(disk)  = complex . 

%Although important for most physical quantities, 
Unfortunately, the distance to L1157 is uncertain.   
The distance of 440 pc, based on spectroscopy and photometry of  
the illuminating star HD200775 of NGC 7023, 
is commonly used \citep[see][]{Kun2008},   
while the distance of the neighboring L1147/L1158 complex was determined to 
be 325 pc by the extinction-distance relationship in \citet{Straizys1992}.  
Later, the Cepheus cloud was found to have three characteristic distances 
200 pc, 300 pc, and 450 pc in \citet{Kun1998}.   
Multiple layers of clouds make it difficult to know the exact distance to a 
specific region. 
For easier comparison with the study of \citet{Looney2007}, 
we adopt 250 pc in this paper.

We present interferometric observations of L1157-mm
using the Combined Array for Research in Millimeter-wave Astronomy 
\citep[CARMA;][]{CARMA} 
\footnote{http://www.mmarray.org}. 
The dust continuum at 3 mm and the N$_2$H$^+$ gas emission are observed
and studied.  The observational setup and the results are presented in \S 2 and
\S 3.  We estimate the column density of the N$_2$H$^+$ emission (\S 4.1)
and compare it with the previous detection of the flattened envelope
at 8 $\mu$m absorption (\S 4.2).  
The N$_2$H$^+$ abundance is examined in \S 4.3. 
% with a brief discussion of chemistry effects 
We construct a simple model to interpret both the dust continuum 
%with expected N$_2$H$^+$ gas emission consistent with the observations 
and the N$_2$H$^+$ gas emission in \S 4.4. 
In addition, the gas kinematics of the large-scale envelope is discussed
in \S 4.5.  An overall link between gas and dust at multiple scales 
in this system is concluded in \S 4.6, following by a summary in \S 5.

\section{Observations} 

The observations were carried out with CARMA  
%on Oct 2nd and Oct 5th 2008 in E configuration. 
in October 2008 in E configuration and March 2009 in D configuration. 
The system temperature ranged from 150 to 300K 
(single sideband, SSB) during source observations. 
D and E configurations are the two most compact configurations of CARMA and 
give angular resolutions of around 5.5\arcsec~and 10\arcsec~at 3 mm, 
respectively.  Because of the short antenna separations (8-66 m) in E
configuration, antennas can shadow each other when observing targets at
low elevations.  To minimize shadowing, the elevations were always above
30\degr~throughout the observations.  Also we checked and verified
that no shadowed data of the source were included in the analysis.

CARMA is composed of nine 6.1m Berkeley Illinois Maryland Association (BIMA) 
antennas and six 10.4m Owens Valley Radio Observatory (OVRO) antennas   
\footnote{Currently eight 3.5 m antennas from Sunyaev-Zeldovich Array (SZA) 
are also combined with CARMA}. 
At our observed frequency, the FWHM of the primary beams are 
122\arcsec~and 70\arcsec~for the 6m and 10m dishes, corresponding to 
30,500AU and 17,500AU, respectively.  
The sensitivity decreases significantly outside the FWHM.  
To observe the extended feature of L1157,  
%with size larger than the FWHM of the antennas, 
we performed a five-pointing mosaic observation across the disk-like structure   
found in \citet{Looney2007}. 
The phase center of the central pointing is set at the position of the 
protostar,  
%L1157-mm, (also known as L1157 IRS or IRAS 20386+6751), 
%$\alpha$ = 20$^h$39$^m$06\fs256 and 
%$\delta$ = 68\degr02\arcmin15\farcs810, 
$\alpha$ = 20$^h$39$^m$06\fs26 and 
$\delta$ = 68\degr02\arcmin15\farcs8, 
determined by the dust continuum peak from other high resolution 
CARMA observations (Chiang et al. in preparation).  
%(from CARMA B-array unpublished observation)

%frequency correlator setup 
The correlator was set so that the N$_2$H$^+$ $JF_1F$=101-012 component 
at 93.1763 GHz was observed with a 2 MHz bandwidth, 
which provided a velocity resolution of 0.098 km~s$^{-1}$.  
Dust continuum was observed simultaneously with two 500 MHz bands. 
During the observations, 1927+739 was observed every 18 minutes and used as 
the phase calibrator.  
%Also, 3C454.3 and 3C84 were bandpass calibrators, while Uranus and MWC349
%were flux calibrators for tracks on Oct 2nd and 5th, respectively.
Also, the bandpass calibrators were 3C454.3, 3C84, 3C345, and 1642+689 for 
different tracks, while the flux calibrators were Uranus and MWC349.  
The absolute flux uncertainty is around 10\%; 
%, which affects the absolute flux but not the relative flux; 
hereafter we only consider the statistical uncertainty.  
All observational data were reduced and imaged using the MIRIAD software 
package \citep{MIRIAD1995}  \footnote{http://carma.astro.umd.edu/miriad}.   
%need a table to summarize the calibrators and source info ? 
For the maps shown in this paper, natural weighting is used 
for Fourier transforming the visibilities into the image space.

\section{Results} 

\subsection{Dust Continuum} 
Figure \ref{figMom0}(a) shows the 3 mm dust continuum map.  
With a beam size of 7.3\arcsec$\times$6.5\arcsec~ (1825 AU$\times$1625 AU)
in the combined D- and E-array CARMA data, the dust emission is compact and
nearly spherical.  We used the MIRIAD task IMFIT to fit the dust emission
with a Gaussian, and the results are given in Table \ref{tabCont}.
We detect the spherical protostellar envelope, but    
%the central protostar and 
the embedded early circumstellar disk is not resolved. 
% at our observational resolution,
The dust continuum seen by CARMA is more compact than that seen 
by single-dish observations 
\citep[e.g., 850 $\mu$m and 1.3 mm maps in ][]{Shirley2000,Gueth2003},   
implying that the large-scale extended emission
%, detected by single-dish telescopes, 
is resolved out by the interferometer.  
A more careful modeling of the dust envelope is done in \S 4.4.

%{\it to-be-deleted -------  
%Assuming optically thin dust at a uniform dust temperature $T_d$ and constant 
%dust opacity, the mass can be estimated by the total flux with 
%\begin{equation}
% M = \frac{F_\nu d^2}{\kappa_\nu B_\nu(T_d)}  
%\end{equation}  
%where $F_\nu$ is the flux, $d$ is the distance to the source, $\kappa_\nu$
%is the mass absorption coefficient, and $B_\nu$ is the Planck function.
%%findflux2 mass=0.354 temp=30 ko=0.0056 freq=91.2 dist=250
%For a simple estimate, assuming $\kappa_\nu$ = 0.0056 cm$^2$g$^{-1}$ at 3 mm 
%and dust temperature 30K \citep{Gueth2003}, the mass is around 0.35 M$_\odot$. 
%%$\kappa_\nu$ = 0.009 cm$^2$g$^{-1}$ at 3 mm (as in Looney2003), 
%% our modeled temperature later is lower, suggesting higher mass 
%Note that interferometry filters out some large-scale emission.  Besides
%the uncertainty of the distance as discussed in the previous section,
%the exact value of $\kappa_\nu$ is very uncertain too because of the
%unclear grain properties.  Also the dust temperature of 30K can be an
%over-simplification too.  
%}

%For example, if considering the coagulated grains in Ossenkopf\&Henning1994, 
%which suggests $\kappa_\nu$ is around 0.005 cm$^2$g$^{-1}$ instead, 
%the simple estimation gives a larger mass of 1.3 M$_\odot$.
%--- this was dust temperature = 10K 
%(coordinate with dust modeling!)

\subsection{N$_2$H$^+$ Gas} 
\begin{figure}
\includegraphics[angle=270,width=0.9\textwidth]{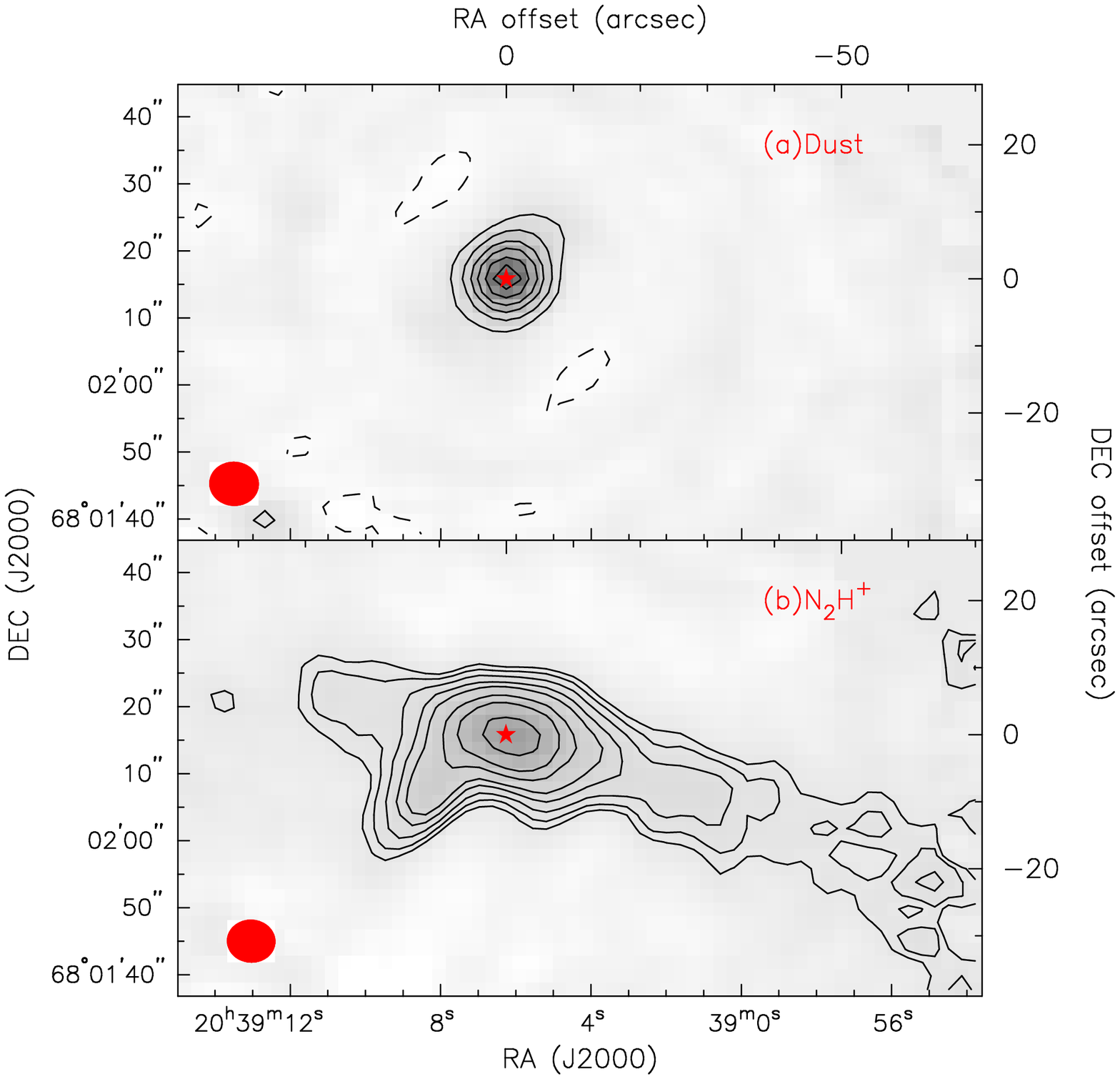}
\caption{
(a) 3 mm dust continuum of L1157 ({\it contours and gray scale}).   
The contour levels are [-2, 3, 6, 10, 14, 20, 26]$\times \sigma$, 
where $\sigma$=1 mJy~beam$^{-1}$ is the noise level, 
and the beam size is 7.3\arcsec$\times$6.5\arcsec~ 
at a position angle of 86.1\degr.  
(b) The integrated intensity of N$_2$H$^+$ over the isolated hyperfine 
component $JF_1F$=101-012 (2.21-3.58 km~s$^{-1}$, 
{\it contours and gray scale}).  
The star marks the position of the central protostar. 
The contour levels are [3, 6, 10, 15, 20, 30, 40, 60, 80]$\times \sigma$,
where $\sigma$=0.01 Jy~beam$^{-1}$~km~s$^{-1}$, 
and the beam size is 7.1\arcsec$\times$6.3\arcsec~
at a position angle of 86.6\degr.  
%(maybe used something to outline the mosaicked region?)
Although negative contours exist due to the filtered-out large-scale
structures, they are not plotted for simplicity.  
%we did not plot them to simplify the map.  
}
\label{figMom0}
\end{figure}

\begin{figure}
\includegraphics[angle=270,width=1.0\textwidth]{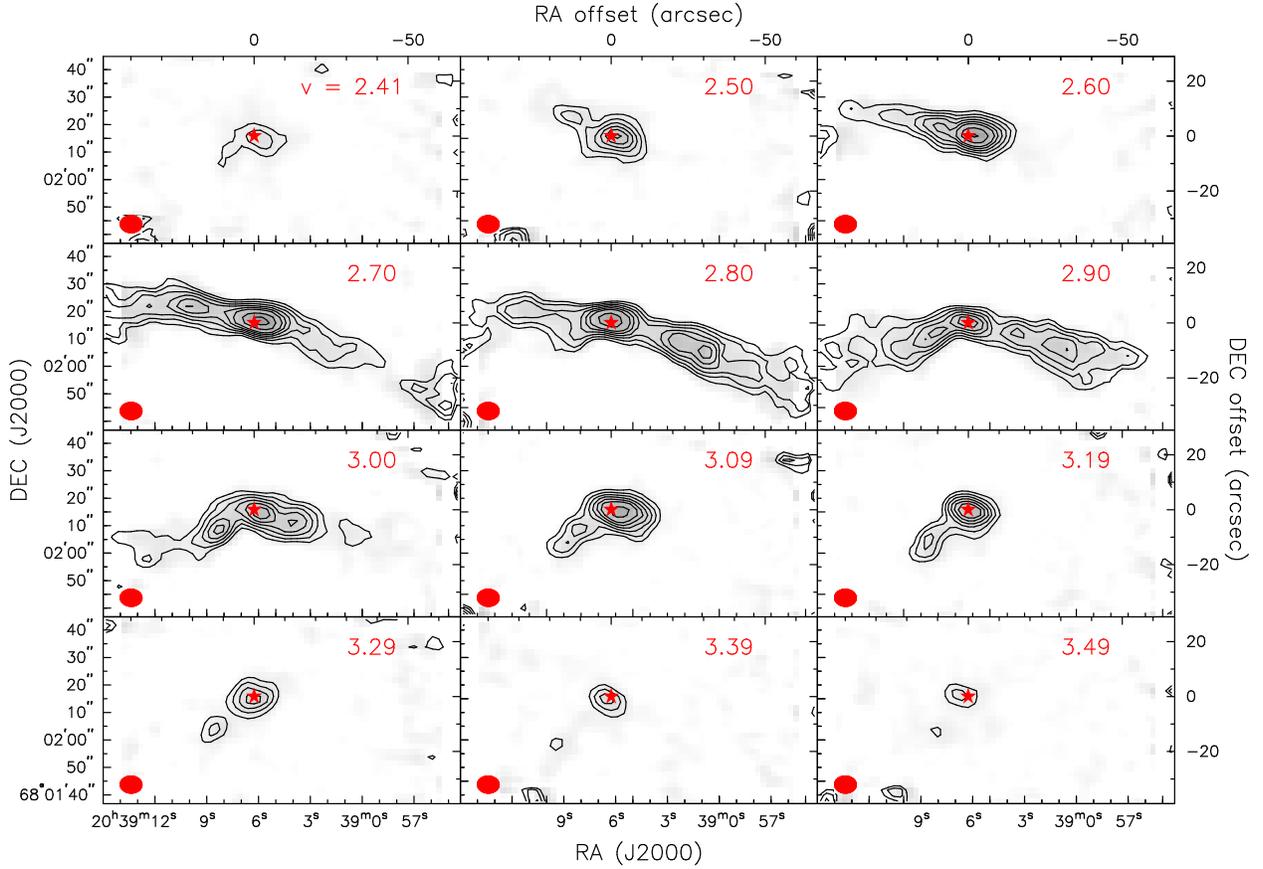}
\caption{ 
Channel maps of L1157 for the N$_2$H$^+$ $JF_1F$=101-012 transition 
({\it contours and gray scale}).   
The numbers in the upper-right corner are the LSR velocity for each channel. 
The star marks the position of the central protostar. 
The contour levels are [2, 3, 4, 5, 6, 7, 9, 11]$\times \sigma$, 
where $\sigma$ =0.08 Jy~beam$^{-1}$,
and the beam size is 7.1\arcsec$\times$6.3\arcsec~  
at a position angle of 86.6\degr.  
Negative contours are not plotted for simplicity.
}
\label{figChanMap}
\end{figure}

Figures \ref{figMom0}(b) and \ref{figChanMap} show  
the integrated intensity (zeroth moment map) and the velocity channel maps   
of the N$_2$H$^+$ isolated hyperfine component $JF_1F$=101-012. 
The star marks the position of the central protostar. 
%, determined by the peak of dust continuum.  
%With a critical density of around 10$^5$cm$^{-3}$ \citep[e.g.,][]{Daniel2006}, 
We detect large-scale N$_2$H$^+$ emission 
across an elongated region perpendicular to the outflow direction  
(see the gray-scale image in Figure \ref{figClnDn}(a) 
for the outflow orientation). 
The disk-like structure extends over 80\arcsec, which corresponds to a 
linear size of $\sim$20,000 AU assuming a distance of 250 pc.
%N$_2$H$^+$ is observed to exist in an elongated disk-like structure 
%perpendicular to the outflow direction. % (as marked in Fig?).  
Due to the large size, it is unlikely to be a rotationally supported disk  
(see \S 4.5.2). 
Hence, we call it a flattened envelope in this paper.  

The N$_2$H$^+$ gas shows interesting dynamics of the flattened envelope 
in Figure \ref{figChanMap}. 
Systematically, the east wing has lower velocity than the west wing, 
suggesting rotation or bulk motion. 
The dynamics of the system will be discussed more in later sections. 
The spatial distribution is clumpy. 
In particular, the extended emission toward the east-south roughly follows 
the interface of the envelope and outflow cavity, suggesting 
dynamical or chemical interaction between outflow and envelope. 
The effects of outflows on the morphology of N$_2$H$^+$ emission are also seen 
in other YSOs \citep[e.g.,][]{Chen2008}. %because CO outflow destroys N2H+ 
In addition, the emission extends to the very west of the observational 
field of view and becomes blended with noise. 
%has another peak blended with noise.  
%To minimize confusion, 
In this study we do not consider the western clumps 
(see Figure \ref{figMom0}(b)); 
%peak in Figure \ref{figMom0}(b); 
instead, we focus on the main part of the flattened envelope.

\section{Analysis and Discussion}  

\subsection{The Column Density of the N$_2$H$^+$ Feature} 

%The column density of the N$_2$H$^+$ gas is estimated using the spectrum.  
%The N$_2$H$^+$ molecule (diazenylium) has 7 sets of hyperfine components from 
%15 J=1-0 rotational transitions; components of each set are degenerate 
%within 10$^{-6}$ Hz \citep[e.g.,][]{Daniel2006}. 
The N$_2$H$^+$ molecule (diazenylium) has 15 $J$=1-0 rotational
transitions; spectroscopically, seven sets of hyperfine components are
observed because the lower state energy levels $J$=0 are degenerate within
10$^{-6}$ Hz \citep[e.g.,][]{Daniel2006}.
The relative rest frequencies of the components were determined by  
\citet{Caselli1995}. 
%(and chem lab work?)
%The $JF_1F$=101-012 component, together with $JF_1F$=101-010 and 101-011   
%at 93.176258 GHz \citep{Lee2001},  
%is not blended with the other sets of hyperfine components, usually 
%called the isolated component, and is ideal for studying dynamics.  
The so-called isolated component, the $JF_1F$=101-012 component together
with $JF_1F$=101-010 and 101-011 at 93.176258 GHz \citep{Lee2001}, is
not blended with the other sets of hyperfine components and is ideal for
studying dynamics.  
We perform spectrum fitting for pixels with more than three S/N detection of
N$_2$H$^+$ on the integrated intensity map, except the very west region
close to the edge of the field of view, where the spectrum fitting fails
to be reliable.
%For simplicity, we assume a uniform excitation temperature $T_{ex}$=
%10 K with an LTE Boltzmann populations of the levels and a constant
%proportionality of intrinsic line strength for hyperfine components with
%the same linewidth.  Also, a Gaussian form of opacity in the frequency
%space are used \citep[cf. GILDAS CLASS HFS procedure and][]{Shirley2005}.
For simplicity, we assume a uniform excitation temperature 
$T_{ex}$ = 10 K with the local thermodynamic equilibrium (LTE) 
approximation for the level populations. 
Also, Gaussian lines with the same line width are used 
(cf. GILDAS CLASS HFS procedure
\footnote{http://www.iram.fr/IRAMFR/GILDAS/}
and \citealt{Shirley2005}).  
The isolated component accounts for a ratio of 3/27 of the total
N$_2$H$^+$ emission \citep{Daniel2006}, with which we convert our observed
data to total emission accordingly.  Then we calculated the best fit 
values of opacity $\langle\tau\rangle$ 
\citep[average $\tau$ among seven hyperfine sets, see][]{Shirley2005},
FWHM of velocity dispersion $\Delta v$, 
and $v_{\mathrm{LSR}}$ for each pixel using the MATLAB function {\it nlinfit}
\footnote{http://www.mathworks.com/}. 
A summary of the spectrum fitting as well as the uncertainty of fitting 
is in Table \ref{tabSpec}.  
%--------------------------------------------
In the optically thin limit and with the LTE assumption, 
the column density can be estimated with  
\citep{Miao1995,Goldsmith1999} 
\begin{equation} 
 \mathcal{N}_{tot} = 2.04 \, 
 \frac{Q(T_{ex}) \, e^{\frac{E_u}{kT_{ex}}}}{ \theta _a \theta _b \nu^3\mu^2 S} 
 \left[ \frac{B_\nu(T_{ex})}{B_\nu(T_{ex}) - B_\nu(T_{bg})} \right]
 C_\tau  \int I_v \, \textrm{d}v \times 10^{20} \textrm{cm}^{-2} \quad , 
\end{equation}   
where 
$\theta_a$ and $\theta_b$ are observational beam size in arcsecond, 
$\nu$ is the frequency in GHz, 
$\mu$ is the dipole moment in debye, 
$S$ is the line strength,   
$E_u$ is the upper state energy level, 
$B_\nu(T)$ is the Planck function at temperature $T$, 
$I_v$ is the specific intensity in Jy~beam$^{-1}$,  %or brightness  
$v$ is velocity in km~s$^{-1}$,
$C_\tau$ is the opacity correction factor
\begin{equation}
 C_\tau = \frac{\tau}{1-e^{-\tau}} \quad , 
\end{equation}
$Q(T_{ex})$ is the rotational partition function 
\begin{equation}
 Q_{rot}(T_{ex}) = \sum^\infty _{J=0} (2J+1) e^{-\frac{E_J}{kT_{ex}}} 
                 \approx \frac{kT_{ex}}{hB}  
\end{equation}
%the partition function can be approximated by  
\citep{Goldsmith1999},  
$E_J = J(J+1) hB$ is the energy level for the rotational transition, 
and $B$ is the rotational constant.  
For linear molecule N$_2$H$^+$, $\mu$ is 3.40 debye and $B$ is 46586.867 MHz  
\citep[JPL catalog:][]{Pickett1998,Green1974}.  
%(or The parameters of N$_2$H$^+$ are listed in Table X?).  
Since the emission is more extended than the beam size, 
the beam filling factor is assumed to be 1 across the map.  
Also, the background temperature $T_{bg}$ is assumed to be constant 2.73K.  

\begin{figure}
\includegraphics[angle=270,width=0.9\textwidth]{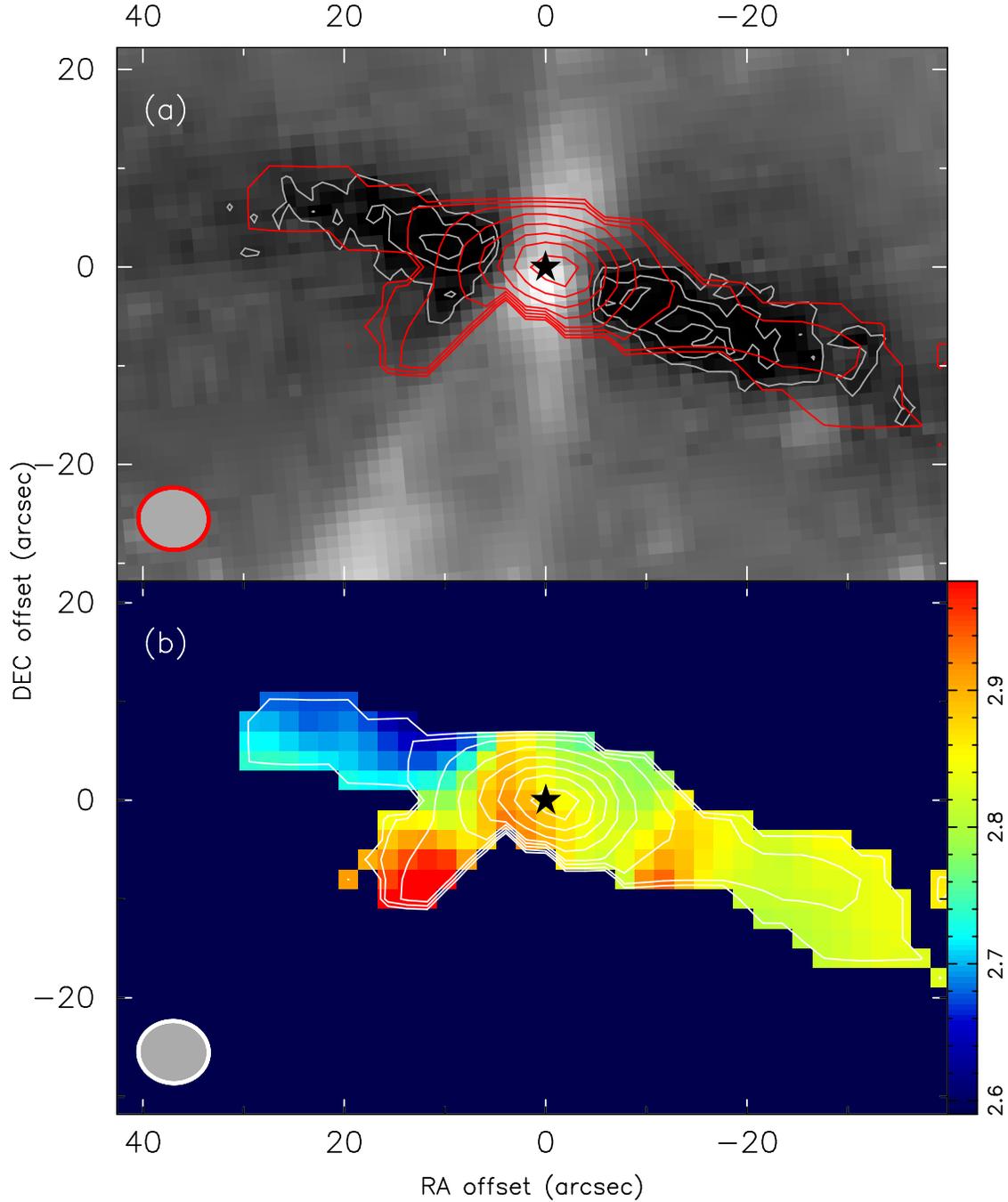}
\caption{ 
(a) IRAC 8 $\mu$m image of L1157 
\citep[][{\it grayscale and gray contours}]{Looney2007} 
overlaid with N$_2$H$^+$ column density ({\it red contours})
derived from the CARMA observations. 
The star marks the position of the protostar. 
The ({\it red}) contour levels for the N$_2$H$^+$ column density 
are [0.1, 0.5, 1.0, 1.5, 2.0, 2.5, 3.0, 3.5]$\times$10$^{13}$cm$^{-2}$ 
(see the text for discussion of uncertainty); and the ({\it gray}) contours 
showing the absorption features are [7, 8, 9, 10]$\times$ $\sigma$,
where $\sigma$ = 0.042 MJy$^{-1}$sr$^{-1}$ is the noise level.  
The circle at the lower-left
corner shows the beam size of the CARMA observations.
(b) Velocity map ($v_{\mathrm{LSR}}$) overlaid by the column density contours from (a), 
both derived from the CARMA observations of the N$_2$H$^+$ flattened envelope. 
}
\label{figClnDn}
\end{figure}

Altogether, the estimated N$_2$H$^+$ column density is shown by the dark 
%({\it black/red}) 
contours in Figure \ref{figClnDn}, 
with the velocity map shown in colormap in Figure \ref{figClnDn}(b).  
%underlaid by 8 $\mu$m image from Spitzer IRAC \citep{Looney2007}. 
The average column density is $\sim$1.0$\times$10$^{13}$cm$^{-2}$ with a    
peak of $\sim$3.8$\times$10$^{13}$cm$^{-2}$ close to the protostar. 
The uncertainty varies across the map and has a mean value of 
$\sim$3$\times$10$^{12}$cm$^{-2}$. % (Table \ref{tabSpec}).  
%the uncertainty is of the order of 10$^{13}$cm$^{-2}$ 
%in the proximity of the central condensation, as discussed in the following. 

%\subsection{(Uncertainty of N$_2$H$^+$ column density estimate)} 
%(justify the assumptions when fitting spectra and estimating the column density... how far can the assumptions fall apart?) 
The uncertainties of the column density estimation are dominated by 
the assumptions of spectrum fitting. 
%(Justify the 10 K Tex assumption) 
First, the largest error source for the derived column density 
is the uncertainty of the excitation temperature $T_{ex}$. 
Typically, the excitation temperature in dense cores and protostars 
is around 4-15K, depending on the source properties 
\citep[e.g.,][]{Benson1998,Shirley2005,Kirk2007,Chen2007}. 
In previous single-dish observations, the N$_2$H$^+$ excitation temperature 
of L1157 was determined to be 8.9K by \citet{Emprechtinger2009}, 
%by spectrum fitting using CLASS,  
in general agreement with the rotational temperature of 
13K obtained in \citet{Bachiller1993} using NH$_3$ emission. 
In this study we assume a constant excitation temperature of 10K, and impose 
an uncertainty of 3K that is propagated with the other parameters.  
We used additional CARMA data that contain all hyperfine components in a
lower spectral resolution for spectrum fitting with the CLASS package,
and confirmed that the excitation temperature is around 10K
near the protostar.

Second, we assume that the N$_2$H$^+$ lines are optically thin. 
%A uniform excitation temperature is more easily reached in this condition as a plus. %(low opacity) 
In fact, the average opacity from our spectrum fitting is 0.31, 
consistent with the assumption of optically thin. 
Nonetheless, the correction factor helps 
mitigate the errors propagated to column density. 
%For example, the opacity correction factor for opacity of 0.34 is 1.18.  
In addition, any self-absorption effect is neglected 
given the optically thin assumption. 
However, the optically thin approximation becomes worse toward the center, 
%where it becomes moderately optically thick with $\tau$=0.63. 
where $\tau$=0.63$\pm$0.03. 

LTE is assumed for the level populations of the 
%hyperfine components as well as the 
rotational transitions.  
%This is a better approximation in high volume density  
%\citep[see a more detail study in][]{Daniel2006}. 
%{\it (boundary density is around $10^5$ to $10^7$!? not sure if i misunderstand Daniel2006) }
%---->seems like a bad reference here?
A constant proportionality between the hyperfine components is also adopted.  
The variation of the relative strengths between the hyperfine components 
as discussed in the appendix of \citet{Shirley2005} is ignored in this study.  

For simplicity, we assume only one velocity component for the spectrum 
fitting, although this is not true in the inner envelope. 
We use a single Gaussian to fit the spectra over the map to estimate the 
N$_2$H$^+$ column density.  The simplification is insufficient 
for the more complex velocity structures near the center, 
% because there are two velocity components, 
but the resulting difference is smaller than the uncertainty caused by 
the excitation temperature assumption. 
% 3.7 vs 3.0 x 10^13 for the central pixel 
For the kinematics of the envelope, see later sections for further discussion. 
 
Unlike most N$_2$H$^+$ studies, we fit the N$_2$H$^+$ spectrum using 
only the isolated component instead of all seven sets of transitions.  
The 2MHz bandwidth in our observations cannot cover all seven sets simultaneously, 
so the excitation temperature distribution cannot be derived based on the 
relative strength between hyperfine components;   
%%We only observed the isolated component so no 
%The relative strength between hyperfine components were not fully obtained   
%to derive excitation temperature distribution over the map;  
%%from solely our observations 
therefore we assume a uniform excitation temperature.  
However, the difference should not be significant; 
it has been suggested that the two approaches give 
results consistent within frequency resolution   
\citep[e.g.,][]{Mardones1997,Emprechtinger2009}. 

Indeed, our derived N$_2$H$^+$ column density of L1157 is 
comparable to other studies.
For example, the column density was reported to be 
1.7$\times$10$^{13}$cm$^{-2}$ in \citet{Bachiller1997} and 
1.1$\times$10$^{13}$cm$^{-2}$ in \citet{Emprechtinger2009},  
both using the IRAM 30m telescope with half-power beamwidth (HPBW) of 27\arcsec~
with different assumptions of excitation temperature.  
%Jorgensen et al. 2005 A\&A 435 177  not clear
Our average column density (1.0$\times$10$^{13}$cm$^{-2}$)
is slightly lower, as expected due to the interferometer resolving out 
large-scale emission.

\subsection{Correlation between N$_2$H$^+$ and 8 $\mu$m Absorption}

The extended feature of N$_2$H$^+$ emission coincides with the flattened 
absorption structure found in \citet{Looney2007}.  
Figure \ref{figClnDn} shows the absorption feature at 8$\mu$m
overlaid with the contours of N$_2$H$^+$ column density.  
At 8 $\mu$m, the polycyclic aromatic hydrocarbon (PAH) emission provides a bright background, against
which the dust extinction is detected.  
Note that the spatial resolution of {\it Spitzer} IRAC (the diffraction
limit is 1.8\arcsec) is better than our N$_2$H$^+$ observations 
%(beamsize is 11.46\arcsec$\times$10.19\arcsec).
(beam size is 7.1\arcsec$\times$6.3\arcsec).
%The absorption structure, representing the dust distribution, is
%well resolved with a pixel size of 1.2\arcsec~ by {\it Spitzer}
%IRAC, while the beamsize of our N$_2$H$^+$ observations is
%11.46\arcsec$\times$10.19\arcsec, which is likely why the 8 $\mu$m
%absorption feature appears to be much more clumpy than N$_2$H$^+$
%in Figure \ref{figClnDn}.
In general, the morphology of the absorbing dust 
is consistent with the N$_2$H$^+$ gas, except the central region where 
the bipolar outflow feature dominates.  
From the 8 $\mu$m map, the observed opacity is obtained given the assumption 
of a constant background estimated off-source.  
In \citet{Looney2007}, a simple model of the flattened envelope was
constructed; however, the power-law index of the density profile
could not be constrained.
%(any density profile implied by 3mm obs?) 

The observed opacities were converted into total column density and mass in 
\citet{Looney2007} with the assumption of a mass absorption coefficient of 
dust plus gas $\kappa_{8.0\mu m}$=5.912 cm$^2$g$^{-1}$ 
from \citet{LiDraine2001}.  The total absorbing mass was estimated to
be 0.19 M$_\odot$ in \citet{Looney2007}.  For a comparison, the mass of
the same extended region is $\sim$0.76 M$_\odot$ based on the derived
N$_2$H$^+$ column density and a hypothetical constant N$_2$H$^+$ abundance
n(N$_2$H$^+$)/n(H$_2$) of 3.0$\times$10$^{-10}$.  Note that 
%both values are underestimates of the total envelope mass, because 
the central part of the flattened envelope is excluded here.   
Although most concentrated, the central region is dominated by the
outflow activities at 8 $\mu$m; therefore the method of optical depth
determination in \citet{Looney2007} is not applicable for regions within 
$\sim$8.4\arcsec~of the protostar.
%total around 0.75 M$_\odot$, 
%Incomplete information of the central part poses an unfair comparison.  
%The estimation in \citet{Looney2007} only considered the absorbing material 
%at 8 $\mu$m, which only includes regions more than 8.4\arcsec~away from the 
%protostar, while the N$_2$H$^+$ feature includes both the extended 
%envelope and the central part. --> so we exclude them out  
Also, the chemistry can become more complicated near the center  
so that the real N$_2$H$^+$ distribution is not trivial. %can be entangled. 
Furthermore, this comparison should not be overemphasized 
because of the following complexities: 
%() the contribution of the central outflow-dominating region, 
(1) an uncertain assumption of the N$_2$H$^+$ abundance, 
(2) interferometric filtering,
(3) the uncertainty of $\kappa_{8.0\mu m}$, and 
(4) ignorance of foreground emission when estimating mass using 8 $\mu$m absorption \citep{Tobin2009}.  
%(e.g., Ragan2009ApJ...698..324R). 
% more about the N$_2$H$^+$ abundance in the next section;  

\begin{figure}
\includegraphics[angle=0,width=1.0\textwidth]{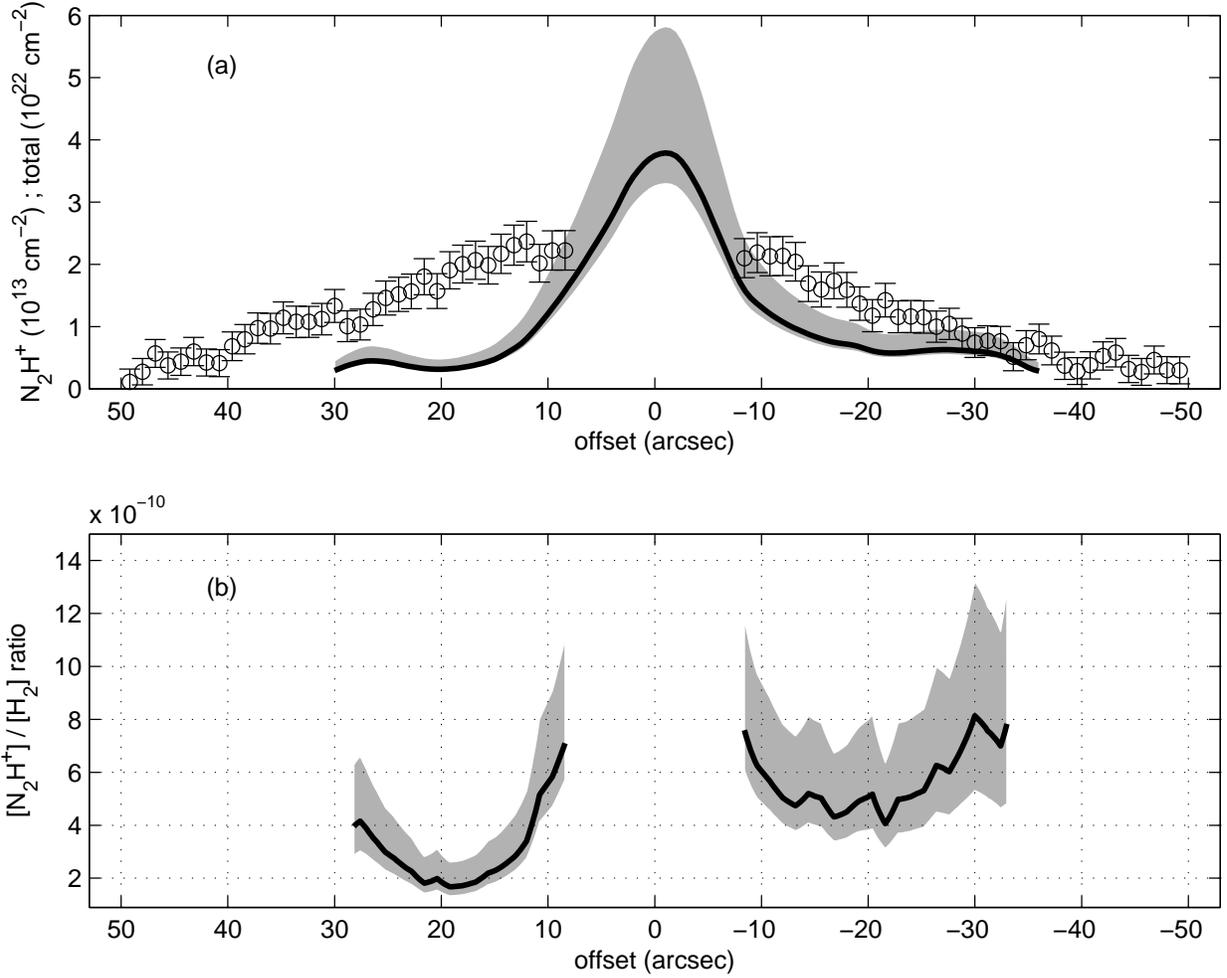}
\caption{ 
(a) Column density of N$_2$H$^+$ (from our CARMA observation, 
{\it curve with shading}) and total gas plus dust (from {\it Spitzer}
observation, {\it circles with error bars}) along the major axis 
of the extended envelope.
(b) N$_2$H$^+$ abundance profile by taking the ratio of [N$_2$H$^+$] 
and [H$_2$] from the CARMA observations and 8 $\mu$m absorption feature.  
The shaded region shows the uncertainty.  %See text for more discussions. 
}
\label{figCut}
\end{figure}

Nevertheless, the total column density, including gas and dust estimated 
from the 8.0$\mu$m extinction in \citet{Looney2007}, 
can be compared with the N$_2$H$^+$ column density 
derived from our spectrum fitting in this study. 
To facilitate the comparison for the flattened structure, we plot the 
profiles along the major axis of the flattened structure in Figure 4.    
%across the disk orientation to see the radial profile of the circumstellar material.
Figure \ref{figCut}(a) shows the N$_2$H$^+$ and estimated total column 
density, derived respectively from CARMA and {\it Spitzer} IRAC observations,  
for a cut of position angle 75\degr~from north to east. 
The circles with error bars are the total column density from the
absorption feature in \citet{Looney2007}, and the thick curves with shades
are the N$_2$H$^+$ column density from this study. 
 
As mentioned in the last section, the dominating uncertainty of the 
N$_2$H$^+$ column density comes from the 
assumption of the excitation temperature T$_{ex}$.  
%As formulated in equation (1)-(3),  %According to equation (1) and (3), 
%the column density is monotonically increasing with and nearly proportional to T$_{ex}$.  
In our spectrum fitting,  a constant T$_{ex}$ of 10K is assumed.
We estimate the uncertainty by imposing an uncertainty of excitation 
temperature $\Delta$T$_{ex}$=3K and the propagated error is shown as a 
color shaded region in Figure \ref{figCut}.  
Because the column density is monotonically increasing with T$_{ex}$  
(from Equations (1) and (3), 
$\mathcal{N}_{tot} \propto Q(T_{ex})e^{\frac{E_1}{kT_{ex}}} 
                   \propto T_{ex}e^{\frac{E_1}{kT_{ex}}}$), 
the upper and lower bounds of the shaded region correspond to T$_{ex}$=13K
and 7K, respectively.

\subsection{N$_2$H$^+$ Abundance}

%A radial dependence of N$_2$H$^+$ abundance is shown in Figure \ref{figCut}(b)
%To better compare N$_2$H$^+$ and the total gas, 
We estimate the N$_2$H$^+$ abundance by taking the ratio of [N$_2$H$^+$], 
from our CARMA observations, and [H$_2$], derived in \citet{Looney2007}, 
along a cut of the extended envelope.  
In Figure \ref{figCut}(b), the ratio of [N$_2$H$^+$] to [H$_2$] 
as a function of offset is plotted. 
Only the region away from the central 
outflow-dominating zone and with more than three S/N of N$_2$H$^+$ detection 
is considered here. 
While the radial profile of the 8 $\mu$m absorption  
shows a good symmetry between east and west wings, 
the profile of N$_2$H$^+$ does not.  The N$_2$H$^+$ column density drops 
in the west wing faster than the east. 
%The average N$_2$H$^+$ abundance is 3.7$\times$10$^{-10}$ in the east wing, 
%around 40\% lower than the average abundance 6.2$\times$10$^{-10}$ in the west wing.  
The average N$_2$H$^+$ abundance is 3.0$\times$10$^{-10}$ in the east wing, 
lower than the average abundance 5.7$\times$10$^{-10}$ in the west wing.  
%compare to other studies. e.g., 
%(abundance 3.8e-9, Bachiller1997) 
%(abundance $>$1.0e-9, Joergensen2004)
Our estimate of N$_2$H$^+$ abundance is smaller than the value derived 
in \citet[][3.8$\times$10$^{-9}$]{Bachiller1997}, 
while the difference can come from a different estimate of H$_2$
column density and interferometric filtering.  
Nonetheless, the overall N$_2$H$^+$ abundance is consistent with
most chemical models.  The typical N$_2$H$^+$ abundance [N$_2$H$^+$]/[H$_2$] 
for prestellar cores and Class 0 YSOs varies from 
10$^{-11}$ to 10$^{-9}$ \citep[e.g.,][]{Aikawa2005,Maret2007B68,Tsamis2008},  
%10$^{-9}$ (e.g., the inner region of Tsamis2008), 
%10$^{-10}$ (Aikawa2005), 
%10$^{-11}$ (Maret2007B68). 
%Our observed N$_2$H$^+$ abundance is slightly higher than the typical values. 
%Besides the uncertainty of the real gas density, 
%it can be due to the evolutionary effect \citep[e.g.,][]{Bergin1997}. 
%%10$^{-10\pm 1}$ (Bergin)
%Maybe pretty consistent with Lee2004? 
and likely changes with YSO evolution \citep[e.g.,][]{Bergin1997}.

Enhancement of N$_2$H$^+$ abundance close to the center is %arguably 
seen in the east wing (Figure \ref{figCut}(b)). 
%, although the trend is not definite given the large beamsize.  
This trend is also expected by chemical models.
While the preceding objects, starless cores, have been shown to have 
constant N$_2$H$^+$ abundance \citep{Tafalla2002}, 
the abundance profile evolves as a function of radius in the collapsing cores 
%For examples, the abundance profiles for various species are
%calculated numerically in 
\citep[e.g., the numerical chemical models:][]{Lee2004,Aikawa2005,Tsamis2008}.
In particular, the formation and destruction of N$_2$H$^+$ is closely related 
to other species by the chemical reactions with N$_2$ and CO  
\citep{Womack1992,Joergensen2004}. 
The dominant route to form N$_2$H$^+$ is through  
\begin{equation} 
 \textrm{H}_3^+ + \textrm{N}_2 \rightarrow 
 \textrm{N}_2\textrm{H}^+ + \textrm{H}_2,  
\end{equation}    
while it is destroyed 
by dissociative electron recombination at high temperature 
\begin{equation} 
 \textrm{N}_2\textrm{H}^+ + e^- \rightarrow \textrm{N}_2 + \textrm{H},  
\end{equation}   
and CO destroys N$_2$H$^+$ in the gas phase by the reaction 
\begin{equation} 
 \textrm{N}_2\textrm{H}^+ + \textrm{CO} \rightarrow 
 \textrm{N}_2 + \textrm{HCO}^+  . 
\end{equation}   
Therefore, when CO starts to deplete onto dust grains 
at densities around 2-6 $\times$10$^4$ cm$^{-3}$ \citep[e.g.,][]{Tafalla2002}  
in the inner envelope, the abundance of N$_2$H$^+$, as well as 
other nitrogen-bearing molecules, increases. 

On the other hand, depletion of N$_2$H$^+$ is also expected 
in the inner envelope.  
%Although harder to deplete due to the low binding energy of its precursor 
%molecules N$_2$, N$_2$H$^+$ is expected to be depleted at density 
%where density is above / exceeds 10$^6$ cm$^{-3}$ or higher such as 3$\times$10$^7$cm$^{-3}$ 
The depletion density is around 10$^6$ cm$^{-3}$ to 3$\times$10$^7$ cm$^{-3}$, 
above which the gaseous N$_2$H$^+$ start to deplete 
\citep{Bergin1997,Aikawa2003}. 
However, whether depletion is seen in our observations is not clear  
due to the fact that the region with density higher than the depletion density 
is not well resolved with the observational beam size. Also, information 
of extinction and total column density is missed owing to the bright 
outflow activity at 8 $\mu$m, making us unable to obtain the N$_2$H$^+$
abundance near the center. 
Furthermore, the CO molecules, existing in the outflows or evaporated from
the dust grains in the very inner envelope due to the heating from the
central protostar, can destroy the N$_2$H$^+$ molecules \citep{Lee2004};
%Furthermore, the CO molecules that exist in the outflows or the very inner 
%envelope due to heating from the central protostar and evaporation from
%the dust grains, can destroy the N$_2$H$^+$ molecules \citep{Lee2004};
the CO outflow and evaporation effects can look similar to the 
depletion effect from the observational point of view.
%Moreover, since the CO molecules in the outflows destroys the N$_2$H$^+$
%molecules, the outflow effect can look similar to the depletion effect
%from the observational point of view.
%%
Eventually, a more careful study for L1157, such as \citet{Evans2005B335} for 
B335, a similar protostar at the same evolutionary stage as L1157, 
will be needed to consider multiple species and  
understand the system more thoroughly.  
%can be interpreted with both step function model and evolutionary chemical 
%model for the chemical profiles 

\subsection{Simple Dust Modeling} 

While a large scale flattened envelope is detected by N$_2$H$^+$ emission, 
the dust emission at 3 mm is compact and round; the goal in this 
section is to construct a model that can interpret both.  
To do so, we begin with fitting the dust continuum. 
We constructed a model that has a flattened geometry similar to the  
N$_2$H$^+$ feature and predicts an observed spherical dust continuum. 

\begin{figure}
\includegraphics[angle=270,width=1.0\textwidth]{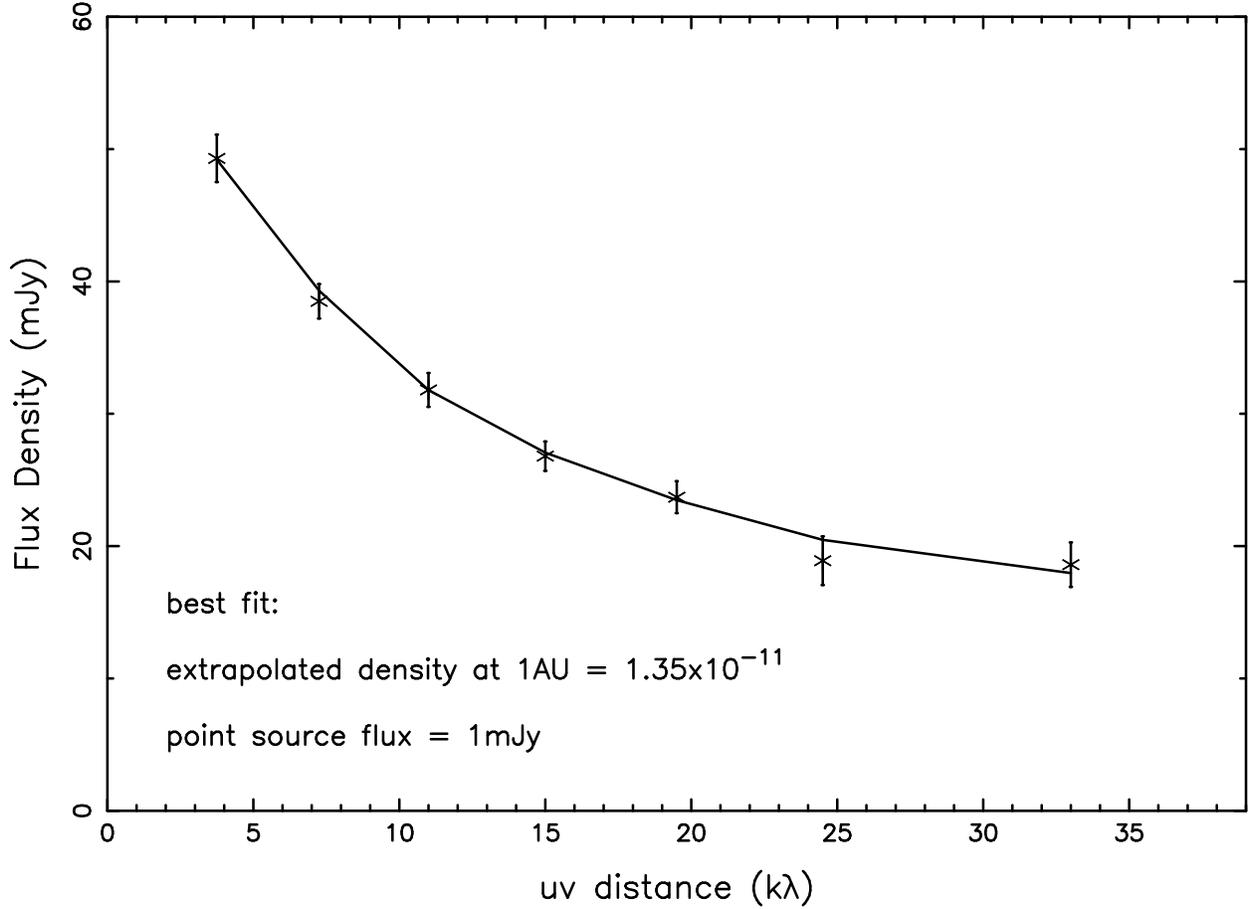}
\caption{
Annuli-averaged flux density for the 3mm dust continuum of L1157 
({\it crosses with error bars}). 
The curve show the best-fit model binned in the same fashion.  
The dust model has a power-law density profile with index = -2 tapered  
by a vertical Gaussian with a scale height of 2000AU.
The extrapolated density at 1AU is 1.35$\times$10$^{-11}$ g~cm$^{-3}$  
and an unresolved point source flux of 1mJy is added. 
The inner radius of the envelope is 5AU, and the outer radius is 15000 AU.
The reduced $\chi_r^2$ is 0.225 for this best-fit model. 
}
\label{figUvamp}
\end{figure}

We have developed a radiative transfer code that considers density and 
temperature structures in three dimensions to do the dust continuum modeling, 
and compare with visibilities from interferometric observations. 
First, a map of flux density is obtained with high numerical resolution  
given a model envelope. 
For each pixel on the plane of sky, the flux is calculated by integrating 
the dust emission along the line of sight \citep[e.g.,][]{Adams1991}.   
%with the consideration of the opacity 
The specific intensity can be expressed as 
\begin{equation}   
 I_\nu = \int_{los} B_\nu(T) \, e^{-\tau_\nu} \, \textrm{d}\tau_\nu 
%       = \int_{los} B_\nu(T) e^{-\tau_\nu} \rho \kappa_\nu d\vec{r},
       = \int_{los} B_\nu(T(\vec{r})) \, e^{-\tau_\nu(\vec{r})} \,  
         \rho(\vec{r}) \, \kappa_\nu \, \textrm{d}\vec{r},
% I_\nu = \int B_\nu(T(\vec{r})) exp(-\tau_\nu) 
% \rho(\vec{r}) \kappa_\nu dl (1-exp(-d\tau_\nu)),
\end{equation}     
%\begin{equation}   
%\tau_\nu(\vec{r}) = \kappa_\nu \int_{los} \rho(\vec{r}) d\vec{r} 
%                  = \kappa_\nu \int_l^\infty \rho(\vec{r}) dl' 
%\end{equation}     
where $B_\nu(T)$ is the Planck function at dust temperature $T$, 
$\tau_\nu$ is the optical depth from the position $\vec{r}$ along the 
line of sight to the observer, 
$\rho$ is the envelope density,  
and $\vec{r}$ denotes the position.  
$T$, $\rho$, and $\tau_\nu$ are all functions of $\vec{r}$. 
%Density $\rho(\vec{r})$ and temperature $T(\vec{r})$ are from envelope model in three dimensions. 
In this study, the opacity of dust grains $\kappa_\nu$ is assumed to be
uniform throughout the envelope.  We simulated all observational
effects from the interferometric observations.
The image is corrected by the primary beam effect according to the antenna
size, and Fourier transformed into visibilities with the observational
{\it u-v} sampling.  In the case of CARMA, the 6m BIMA dishes and 10m OVRO
dishes give three different primary beams.  Therefore we constructed separate
primary-beam-corrected images for each kind of baseline, 
and sampled them with corresponding {\it u-v} spacings from real observations. 
However, the mosaic pattern was not considered in the dust modeling, 
although a five-pointing mosaic was used in the observations.  
%to cover the extended feature of N$_2$H$^+$. 
The dust continuum is very compact with size much smaller than the 
FWHM of the primary beam, hence the effect is insignificant. 
Finally, the modeled visibilities are binned into {\it u-v} annuli around 
the source center and averaged vectorially.  The flux density for each bin 
is compared with the observations, which are binned in the same way 
using the MIRIAD task UVAMP, as shown in Figure \ref{figUvamp} 
by crosses with error bars. 
A $\chi_r^2$ value is calculated to quantify 
the goodness of a fit.  The comparison is done in visibility space 
so uncertainty from imaging processes such as the CLEAN algorithm is avoided.  

The code has been tested for accuracy and convergency. 
The results of spherically symmetric cases 
were checked by comparing with those obtained by 
previous codes used in \citet{Looney2003} and  \citet{Chiang2008}. 
Because of the nature of higher density and higher derivatives 
in the inner envelope, smaller step sizes are required for the inner region.  
On the other hand, step size is not as sensitive for the outer region,
so computing time is saved by using coarser steps.  In the current code,
three refinements of step sizes are used with extra fine steps taken for
the very central region to achieve accuracy better than observational
uncertainties.  Also, cases of different numerical resolutions have been
run to check convergency.  In this study, the numerical resolution is
37.5AU, corresponding to 0.15\arcsec~for L1157.  Higher numerical
resolution gives consistent flux density within a difference smaller
than 0.05 mJy.  In addition, the code itself does not assume any symmetry.
In other words, the model of envelope density and temperature can be arbitrary 
in three dimensions. 
%(quantitative estimate such as accuracy vs stepsize etc. 
% it was stupid... the future version will have adaptive stepsize!) 

With the observed flattened structure in mind, we construct a model envelope  
of similar morphology. 
For simplicity, we adopted a power-law density and temperature profile. 
To impose the flattened geometry, we use a spherically symmetric  
distribution and taper the density profile with a Gaussian mask 
diminishing along the minor axis of the disk orientation. 
%perpendicular to the disk extension direction. 
%It is different from the hydrostatic equilibrium disk since it still carries 
%the power-law properties, just the spherical symmetry is broken. 
In our model, the vertical scale height of the Gaussian is fixed to be 
2000 AU for an ellipticity close to the observed flattened feature. 
We orientate the flattened envelope as the observed N$_2$H$^+$ feature, 
that is, with a position angle of 75\degr. 
We also adopt an inclination angle of 80\degr,   
%with the south facing toward the earth, 
as determined from outflow observations \citep{Gueth1996}. 
However, the temperature distribution remains spherically symmetric. 

%{\it (looking for similar references...)  
%(more justifications on density temperature geometry ...)}
Note that this model is not physically motivated, 
but it interprets the observations with statistical significance.  
We adopt a power-law index of -2 for the density structure, 
similar to a singular isothermal sphere in the extreme case of 
a Bonnor-Ebert sphere \citep{Shu1977}. 
%as the outer region of the BE sphere that fits prestellar cores
%observationally (e.g. KirkWard-Thompson2005) and expected from theoretical
%calculations.  The power law of -2 density profile is established in
%some other theoretical calculations as well \citep[e.g.,][?]{Gomez2007}.
The dust temperature profile is a power law of index -2/(4+$\beta$),
where $\beta$ is the spectral index of opacity, from the spherically
symmetric approximation assuming the heating is dominated by the central
source \citep{Adams1991}.
We set the extrapolated temperature at 1AU to be 300K and a lower limit of 5K 
representing the external heating from cosmic ray or interstellar radiation. 
%{\it 
%We also tried density profile with power-law index of 1.9 or smaller,
%they give fewer good fits...  physically power law index should not
%be larger than 2.0, SIS,   also because there is plenty of mass supply
%surrounding (right?)  (read VorobyovBasu2006) 
%We also tried a power-law index 1.5, implying the free-fall collapse is 
%undergoing. Heavier central point-source is required in this scenario.
%}

For the dust grain properties, 
a fiducial value of the mass opacity coefficient 
$\kappa_{3mm}$=0.0056 cm$^2$g$^{-1}$ is adopted, 
based on $\kappa_{1mm}$=0.01 cm$^2$g$^{-1}$ \citep[e.g.,][]{Natta2004} 
and $\beta$ = 0.5 \citep{Kwon2009}. 
The exact value of $\kappa$ is uncertain and can vary by an order of 
magnitude dependent on the grain properties 
(e.g., the interstellar MRN grains in \citealt{DraineLee1984} 
and the coagulated grains in \citealt{OH1994}).  
Nevertheless, $\kappa_\nu$ is expected to follow a power law with 
frequency $\nu$ at long wavelengths, that is, $\kappa \propto \nu ^\beta$.   
The extrapolation at millimeter wavelengths is determined by the 
spectral index $\beta$, which is dependent on grain size, grain geometry,
chemical composition, etc \citep{Draine2006}.  For a young object
like L1157, grain growth may have already started in the inner region.
For simplicity, the dust grain properties is assumed to be uniform across
the envelope and not dependent on the radius.
The possible radial gradient of grain properties,
which was recently reported by \citet{Kwon2009}, is neglected in this study.  

\begin{figure}
\includegraphics[angle=0,width=1.0\textwidth]{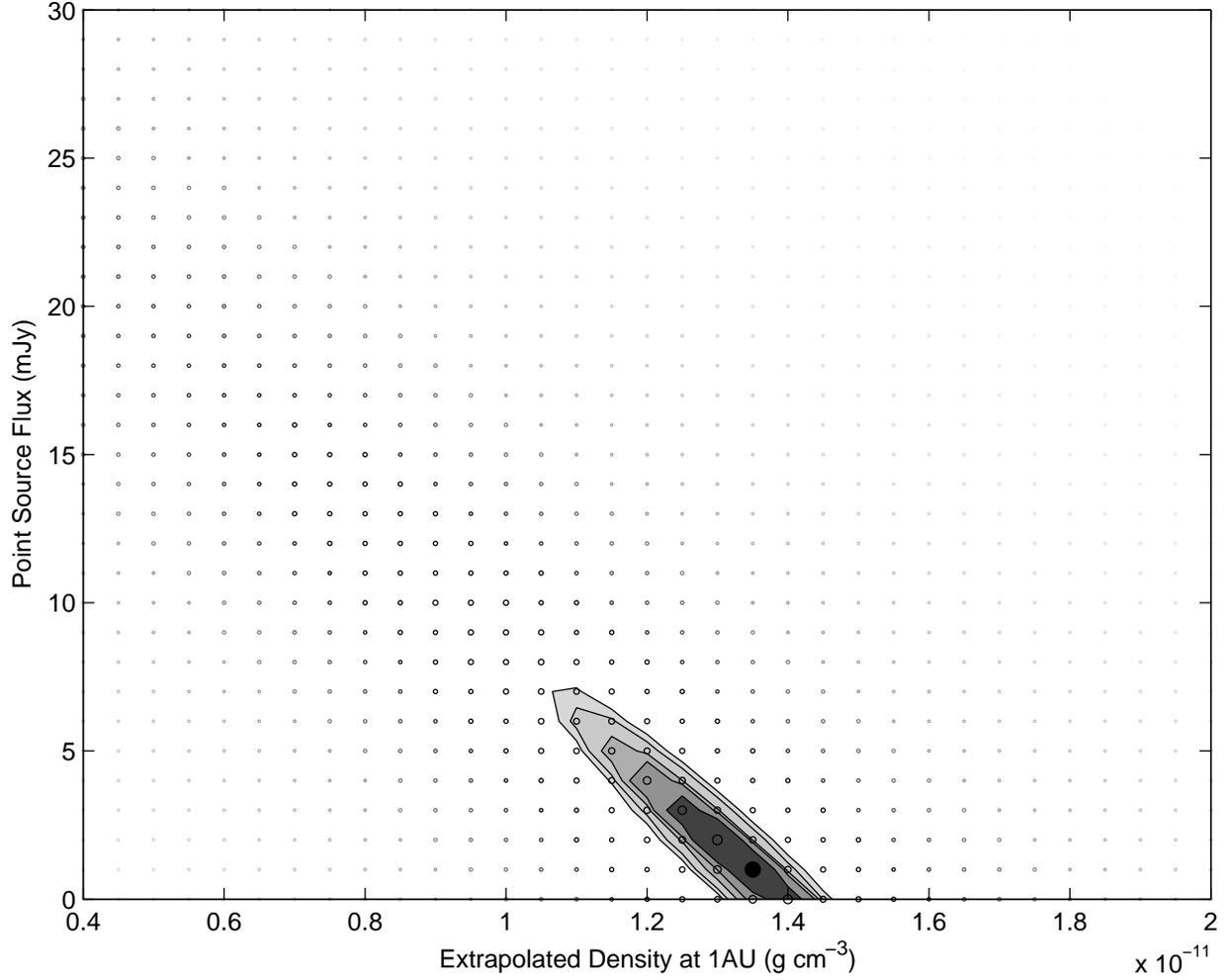}
\caption{ 
Summary of the dust continuum fits. 
Circles show the grid of models with sizes and shades 
representing the goodness of fits. 
Contours are confidence levels of 99\%, 95\%, 90\%, 80\%, and 50\%.  
The filled circle at 1.35$\times$10$^{-11}$g~cm$^{-3}$ and 1 mJy  
is the best fit. 
%illustrate 
}
\label{figFits}
\end{figure}

%(parameter search, chi-square, grids) 
We have explored two parameters:  
point-source flux ($F_p$), representing the flux contribution from the embedded 
circumstellar disk system inside the inner radius of the model envelope, 
and the extrapolated density at 1AU ($d_1$), scaling with the total mass. 
%We searched through the parameter space by doing every possible combination of parameters.  
A grid of model parameters is run with $F_p$ from 0 to 30 mJy
in steps of 1 mJy, and 
%$d_1$ from 4$\times$10$^{-12}$ to
%16$\times$10$^{-12}$g~cm$^{-3}$ in steps of 1$\times$10$^{-12}$g~cm$^{-3}$.   
$d_1$ from 4$\times$10$^{-12}$ to
20$\times$10$^{-12}$g~cm$^{-3}$ in steps of 0.5$\times$10$^{-12}$g~cm$^{-3}$.   
%while all other model properties are fixed. 
The inner and outer radius of the envelope are fixed to be  
5AU and 15,000AU, respectively.  
Varying the outer radius of the envelope does not make 
a big difference because little flux is emitted from the low-density region 
compared to the denser region. 
Also, the interferometer resolves out some of the large-scale emission. 
The inner radius is correlated to the point-source flux and can be chosen 
as long as it is much smaller than the observational spatial resolution, 
%while physically there has to be a cutoff radius of the envelope.  
while physically the dust destruction radius is of the order of 0.1AU, 
inside which the temperature is too high for dust to exist. 
% dust destruction radius from Hartmann book page 62 
%inside which the protostar and accretion disk take over.  
%Nonetheless, the presumably 
%embedded circumstellar disk is not resolved by our observations.  
%A $\chi^2$ value is calculated for each set of parameters.
A summary of the $\chi_r^2$ model fitting is shown in Figure \ref{figFits}.  
The filled circle marks the best fit and the contours show the parameter 
ranges with different confidence levels. 
The two parameters are correlated and cannot be clearly distinguished, 
because either a big density scaling factor $d_1$ 
or a high point-source flux $F_p$ can result in a high peak of flux density. 
%best fit for plaw2 z2000
%E array 1.300E-11    5.0   15000.0  0.500    0.003   0.554 
%D+E     1.350E-11    5.0   15000.0  0.500    0.001   0.225
The best-fit model has a point-source flux $F_p$ = 1 mJy and extrapolated 
density at 1AU $d_1$ = 1.35$\times$10$^{-11}$g~cm$^{-3}$. 
A total envelope mass of 1.5 M$_\odot$ is then implied in our model;  
%plus an unresolved mass from the embedded disk-protostar system, 
%comparable to the 
this mass is generally consistent with other mass estimates
%envelope mass of 2.1 M$_\odot$ estimated in \citet{Gueth2003} 
\citep[e.g.,][]{Shirley2000,Gueth2003,Froebrich2005}. 
%the mass estimated from the protostar's luminosity 
The modeled flux density of the best fits as a function of $uv$ distance
is shown by the curve in Figure \ref{figUvamp}. 

\begin{figure}
\includegraphics[angle=270,width=1.0\textwidth]{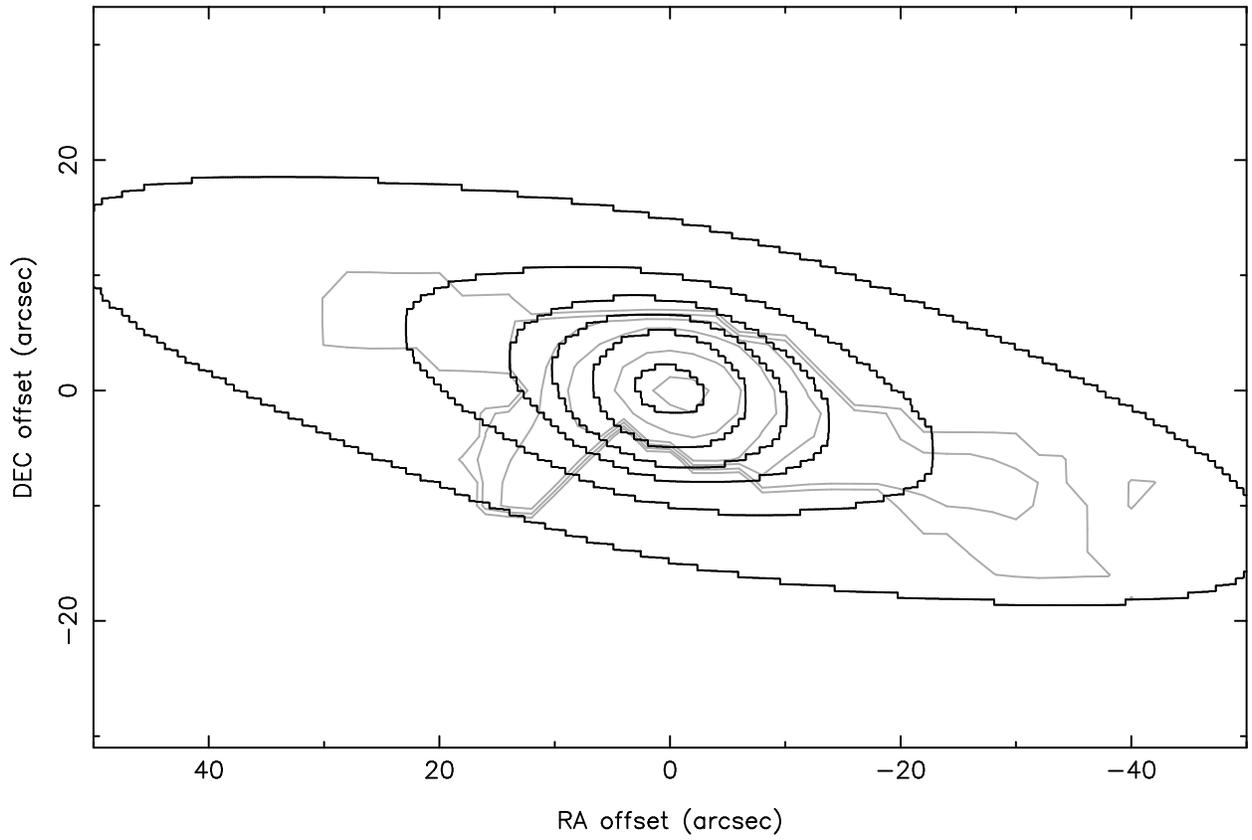}
\caption{ 
N$_2$H$^+$ column density of the best-fit model by dust continuum modeling 
({\it black contours}) and observations ({\it grey contours}).  
The contour levels (for both) are 
[0.1, 0.5, 1.0, 1.5, 2.5, 3.5]$\times$10$^{13}$cm$^{-2}$.
The depletion effect is taken into account at the central region 
where the density is above 1.5$\times$10$^6$cm$^{-3}$.
}
\label{figModCD}
\end{figure}

We estimate the corresponding N$_2$H$^+$ column density of the best-fit 
model obtained from the dust continuum fitting. 
To do the conversion, the dust-to-gas ratio is assumed to be uniformly 1/100, 
as the typical value in the interstellar medium, and the N$_2$H$^+$ abundance  
is assumed to be uniformly 3.0$\times$10$^{-10}$.  
Further, we simulate the depletion effect by introducing a threshold 
depletion density, 
%, above which the N$_2$H$^+$ molecules start to be depleted onto grains and 
above which the N$_2$H$^+$ density does not increase with H$_2$ density.  
We adopt a depletion density of 1.5$\times$10$^6$cm$^{-3}$.  
%from \citet{Bergin1997}.
To compare with observations, N$_2$H$^+$ volume density is integrated
along the line of sight to calculate the column density across the map.
The model N$_2$H$^+$ column density and the detected N$_2$H$^+$ emission
are shown in Figure \ref{figModCD}.  

As seen in Figure \ref{figModCD}, our best-fit dust model gives a  
consistent map of N$_2$H$^+$ column density.    
%Why don't we see extended dust continuum as the extended N$_2$H$^+$ feature?
This simple model illustrates how we do not observe extended dust continuum 
while the N$_2$H$^+$ emission is detected to be extended more than 
80\arcsec~across. 
%The density and temperature at the outer region of the flattened envelope 
%are too low to generate flux that can be detected by this observation. 
The dust emission, which depends on both density and temperature, 
in the outer region of the flattened envelope is too 
weak to be detectable by these observations.

\subsection{Gas Kinematics}
\begin{figure}
\includegraphics[angle=270,width=0.8\textwidth]{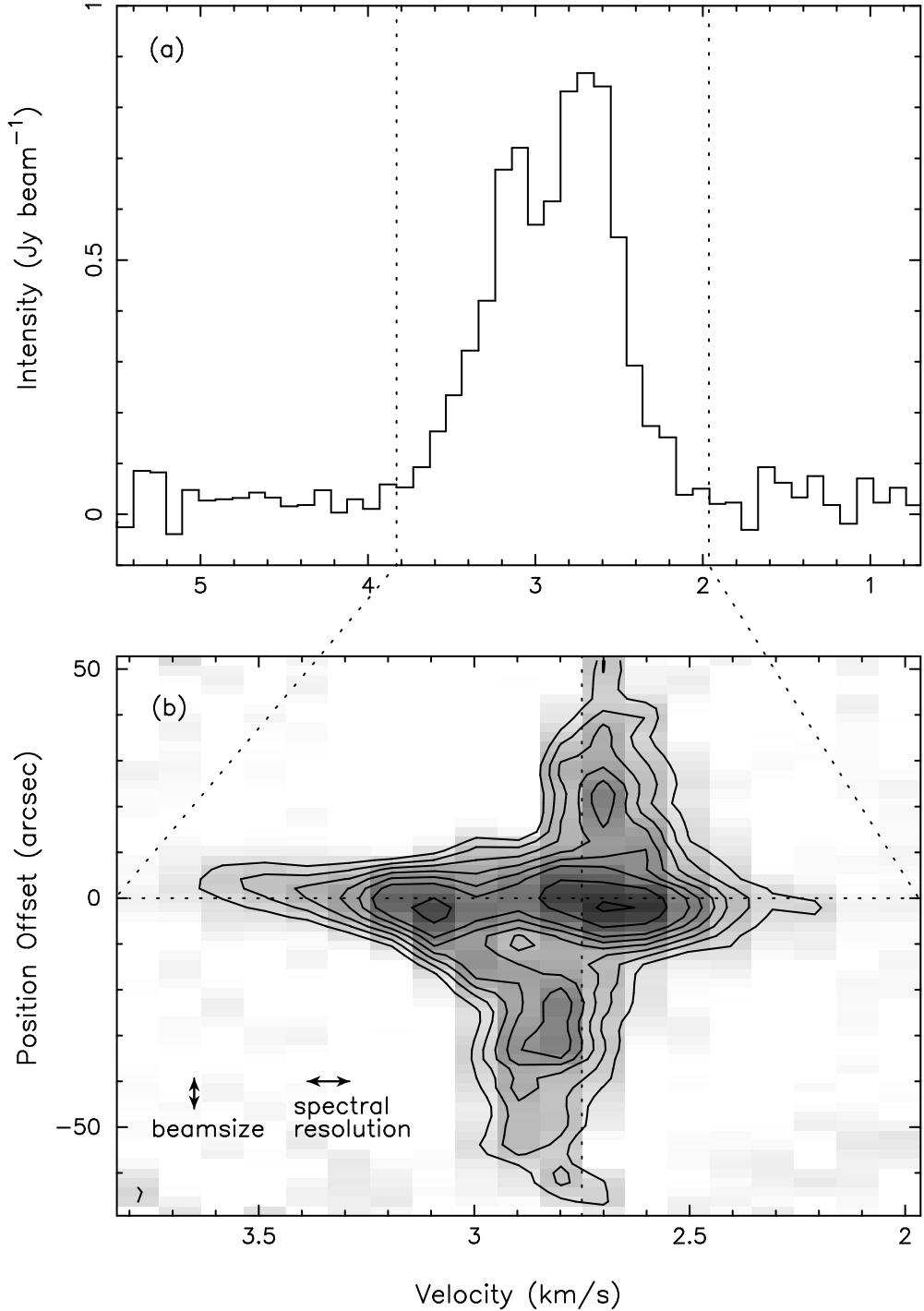}
\caption{ 
(a) Spectrum of the N$_2$H$^+$ $JF_1F$=101-012 line 
at the location of the protostar. 
(b) Position--velocity diagram of the N$_2$H$^+$ $JF_1F$=101-012 line  
along the major axis of the flattened envelope feature, 
a slice with position angle of 75$\degr$.  
%Contours start at 2$\sigma$ and increase in steps of 1$\sigma$  
%($\sigma$ = 0.2 Jy~beam$^{-1}$). %(0.17 )
The contour levels are [2, 3, 4, 5, 6, 7, 9, 11]$\times \sigma$, 
where $\sigma$ =0.08 Jy~beam$^{-1}$.
The angular and spectral resolutions of observations are shown, 
and the pixel size of the gray scale %for the offset 
is the interferometric imaging cell size. 
%The curves represent projected Keplerian rotation with an inclination
%angle of 80$\degr$.  The thick curve is for M$_\star$=0.02 M$_\odot$,
%while the thin curves are for M$_\star$=0.01 and 0.03 M$_\odot$.
%(maybe donot plot the Keplerian curves? since Keplerian rotation is not suggested)
}
\label{figPV}
\end{figure}

The complex N$_2$H$^+$ spectral information reveals a composite 
system of multiple dynamic components (Figure \ref{figClnDn}(b)).  
To understand the detailed kinematics, we perform a simple analysis 
using position-velocity (PV) diagrams.  
Figure \ref{figPV}(b) shows the PV diagram for the N$_2$H$^+$ 
$JF_1F$=101-012 component along the major axis of the flattened envelope 
%disk orientation 
at position angle 75\degr~from north to east. 
This transition is isolated so no confusion is caused 
due to blending with other hyperfine lines. 
%In the following we discuss each feature seen in the position-velocity diagram

We construct a simple model to simulate the PV diagram assuming various
velocity structures of the envelope.  Note that the kinematics modeling
is independent of the dust modeling, except that the best-fit density
structure from the previous section is used.   
%, assuming the same density profile from the best fitted dust modeling 
%A protostellar envelope of the same properties as obtained in the dust modeling is considered.  
We adopt the procedure as discussed in \citet{Ohashi1997} 
and consider a spatially thin cut along the major axis of the envelope.
%In other words, the thickness effect 
Both the inclination and opacity effects are ignored.  
%and the emission is assumed to be optically thin. 
%that is, no opacity effect is considered. 
%For given velocity and density structure of the model envelope, 
%For a given velocity structure of the model envelope, 
The column density in each velocity channel as a function of offset is
estimated by integrating density along the line of sight and sorting into
observational spectral bins by the projected velocity.  The depletion
effect is simulated by assuming constant N$_2$H$^+$ density
for the central region where H$_2$ gas density is higher than 
1.5$\times$10$^6$cm$^{-3}$, 
the same depletion density as adopted in the previous section. 
With the depletion effect, the actual density structure of the inner 
envelope does not play a large role.  
%We simply use constant N$_2$H$^+$ density 
%once depletion takes over in the inner envelope.   
The spatial distribution is then convolved with the observational beam. 
However, unlike the dust modeling, the interferometric filtering effects are 
not taken into account for the kinematics study.  
%Although interferometric effects are not taken into account in the PV modeling,
Nonetheless the large-scale emission from background clouds is not included; 
this is similar to interferometry resolving out large scale emission.   
Finally, the modeled results are shown by contours in PV 
diagrams and compared with observations.

\subsubsection{Spectroscopic Signatures of Infall Motion}

The double-peaked feature seen in the central region 
indicates multiple velocity components in the protostellar envelope  
(Figure \ref{figPV}(a)).  
The redshifted and blueshifted peaks have
a velocity difference of $\sim$0.4~km~s$^{-1}$, inferring a relative
motion between two parts of the envelope.
One interpretation for this is the infall motion of the inner envelope, 
which can also be the cause of the high-velocity wings. 
Gravitational collapse takes place in the early stage of star formation 
\citep[e.g., theoretical studies in ][]{Shu1977,Hunter1977,Tassis2005a}. 
%Besides, accretion disks is expected by the outflow model 
%too \citep[e.g., reviews by ][]{Bachiller1996}. 
However, the predicted infall velocity structures are very different 
from model to model, especially in the inner envelope. 
Detailed comparison of various theoretical models with observations is beyond 
the scope of this paper and may be studied with observations  
of higher spatial and spectral resolutions in the future.  
Nevertheless, infall is expected in the inner envelope of L1157. 
Previously, L1157 has been identified as an infall candidate by 
single dish \citep{Gregersen1997,Mardones1997} 
and interferometric observations \citep{Gueth1997,Beltran2004}.
In addition, detection of methanol from the accretion shocks on the small scale 
also supports the picture of envelope material infalling 
onto the forming protostellar disk \citep{Velusamy2002}.  

Figure \ref{figPVMS}(a) shows the modeled PV contours for a pure infall
motion.  A simple velocity profile $v \sim r^{-1/2}$, representing ideal
free-fall, is used.  
The simple infall model shows a double-peaked feature toward the center
of the envelope, as obtained by \citet{Ohashi1997}.  
However, the model cannot explain the asymmetry of the peaks.
A more sophisticated model that considers full geometry, 
radiative transfer, and different theoretical profiles 
such as \citet{Momose1998} may be necessary. 
%A finite infall region gives rise to the triple peaked feature (?), 
%while the central velocity peak is mainly contributed by the static envelope outside the infall radius.  

We exclude this double-peaked feature as arising from self-absorption, 
commonly used as an indicator of infall motion in the prestellar cores   
\citep[e.g.,][]{Evans1999}. 
In such an approach of identifying infall candidates, a stronger 
blueshifted peak and a weaker redshifted peak are expected 
for infalling sources observed by an opaque line    
because the front half of the cloud causes a redshifted absorption dip 
\cite[e.g.,][]{Myers1996,Masunaga2000},  
while an optically thin line should be observed within two velocity peaks. 
In this specific case, since the N$_2$H$^+$ lines are moderately optically
thin, this scenario of self-absorption is ruled out, although a similar
profile is observed.
%The small but not negligible opacity of N$_2$H$^+$ lines can also cause 
%the blue-skewed asymmetry.  
However, the spectrum is sensitive to the detailed
envelope model such as optical depth, turbulent velocity dispersion,
spatial structures of velocity, density, and excitation temperature,
etc \citep[e.g.,][]{Zhou1992,Ward-Thompson2001}.
The complexity makes it difficult to justify the real physical 
properties causing the observed spectra.  
In addition, the interferometry selectively observes the small-scale
emission, leaving out the large-scale static cloud in the outer envelope
and resulting in spectrum with a deeper zero-velocity dip \citep{Choi2002}.

While a double peaked profile of N$_2$H$^+$ %with a narrow linewidth  
is seen in our interferometric observations, \citet{Mardones1997} 
observed the same line using IRAM 30m telescope, and found 
it to be single-peaked %with a wide linewidth 
and used it as the optically thin 
reference to compare with other optically thick lines for studying infall.  
The same source was observed again by the IRAM 30m telescope in 
\citet{Emprechtinger2009} and similar results were obtained.   
\footnote{However, our recent spectral data taken by IRAM 30m seems
to show a double-peaked feature in the strongest hyperfine component
(J.J. Tobin et al. in preparation). More careful examination is in process.  }  
These support our spectrum fitting results that the N$_2$H$^+$ J=1-0
lines are optically thin.
The major differences of our observation are 
a smaller beam size and the aperture synthesis.   
Therefore the cause of discrepancy is deduced to be 
(1) a beam-smearing effect and/or (2) interferometric filtering. 
While the overall spectrum is dominated by the inner envelope, 
the small-scale structures are not distinguishable by   
the single-dish observations. 
However, observations of higher resolutions with interferometers 
reveal other complexities. 
For example, the degree of profile asymmetry can be enhanced
if observed with a higher angular resolution \citep{Choi2002}.  
Also, we test our modeling routine by convolving the spectrum
with different beam sizes, and confirm the beam-smearing effect.
A double-peaked feature can look like a single Gaussian if the
observational beam size is not able to resolve the infall radius.
%{\it (But Wilner2000 has it while infall radius is smaller than obs resolution!?) }

Second and more importantly, 
interferometric observations reveal only the targeted protostellar envelope  
at the expense of missing flux from large-scale structures, 
arguably dominated by the static foreground and background clouds. 
%The selective detection makes it difficult to compare with single-dish data. 
\citet{Choi2002} has shown that missing short spacing flux can affect the 
self-absorption dip for optically thick lines.  
It can cause similar effects for optically thin lines too. 
In particular, large scale background material is mostly static compared 
to the infalling inner envelope and contributes more flux at the LSR velocity. 
%Star formation is expected to happen in the clouds anyway.  
Nonetheless, the material in the large scale cloud does not participate  
in star formation as actively as the inner envelope.
For comparison with the spectrum of IRAM 30m single-dish observations   
\citep{Mardones1997,Bachiller1997,Emprechtinger2009},  
we smooth our images with a 27\arcsec~Gaussian beam 
(the beam size of IRAM 30m observations) 
%We compared our spectrum with IRAM 30m single-dish observations 
%by averaging our higher resolution image with their beamsize (27\arcsec),  
%CARMA E-array only  4.9404K km/s 
%CARMA DE 7.6373 
%IRAM30m 8.38 (Emprechtinger) 
and found that $\sim$20\% of the total flux is filtered out by CARMA, 
also causing a dip around the LSR velocity.

A large line width of N$_2$H$^+$ J=1-0 is found by the single dish observations. 
%, while the spectrum of our CARMA observation is much narrower.  
The fitted line widths of the single-peaked spectra are 
0.71 and 0.65~km~s$^{-1}$ for the IRAM 30m observations in 
\citet{Mardones1997} and \citet{Emprechtinger2009}, respectively. 
We compare the single-dish observations with our smoothed spectrum. 
If we fit our double-peaked spectrum with a single Gaussian, 
%to represent two velocity components, 
%the FWHM is 0.76$\pm$0.02~km~s$^{-1}$ for the central pixel.  
%the FWHM is 0.74$\pm$0.03~km~s$^{-1}$ for the central pixel.  
the FWHM is 0.88$\pm$0.05~km~s$^{-1}$. % for the smoothed spectrum. 
But if we fit the spectrum with two Gaussians of the same width, the FWHM is  
%0.37 $\pm$0.02~km~s$^{-1}$, comparable to the 
0.44 $\pm$0.02~km~s$^{-1}$,  
comparable to the sonic line width $\sim$0.45~km~s$^{-1}$ and 
larger than the thermal line width $\sim$0.13~km~s$^{-1}$ with the assumed 
temperature of 10K.   
%(maybe show and fit our central spectrum if different sizes of beam are convolved... )
The dynamics at small scales can be obscured in single-dish observations 
but can be revealed by high-resolution interferometric observations.  
If this phenomenon is common to other sources, 
it may be why the average line width of Class 0 YSOs is large 
\citep[$\sim$0.61~km~s$^{-1}$,][]{Emprechtinger2009} compared to 
that of starless cores \citep[0.2-0.4~km~s$^{-1}$,][]{Lee2001}.  
%(Crapsi2005, 0.26km/s ?). 
%%%%In addition, turbulence is at a subsonic level, if present. 
%The non-thermal linewidth dominates given active kinematic motion. 
%compared to turbulence velocity sound speed 
%usually wide linewidth (of order 0.2-0.3 km/s) 
%e.g. starless cores survey LeeMyersTafalla 2001 (look for better references for survey) 
%dominated by thermal instead of turbulence (Myers\&Bensen 1983 or 1989)

%other possible scenarios...
We cannot exclude the possibility that the double-peaked feature is caused 
by factors other than infall motion in the inner envelope. 
For example, outflows contaminate the molecular tracers of envelope material. 
A similar case of the Class 0 protostar B335 was studied by
\citet{Wilner2000}.  While the single-dish observation matched well with
the inside-out infall model \citep{Zhou1993}, the interferometric
observation brought up a more realistic scenario.  
The optically thin CS J=5-4 line 
%with no traits of self-absorption
is shown to be dominated by small-scale outflow clumps in high-resolution
observations \citep{Wilner2000}.  
A single-peaked spectrum seen in low-resolution observations can contain 
two velocity components and hence a double-peaked feature is shown with  
interferometric filtering.
However, some molecular lines do not trace outflows as closely as other
species.  It has been suggested that N$_2$H$^+$ lines trace the quiescent
cores but not the shocked outflow gas \citep{Bachiller1996}.
The CO molecules in the outflows can destroy the N$_2$H$^+$ molecules 
(Equation (6)).  
Indeed, while the abundance of some species is enhanced by the outflow shocks, 
N$_2$H$^+$ is not detected at the shock regions \citep{Bachiller1997}.  
In other words, the outflow contamination is minimized for this specie, 
although some outflow effects are unavoidable (for example, see
the outflow-envelope interface at the southeast extension in Figure
\ref{figMom0}(b)).

The double-peaked spectrum can also be caused by unrelated dense clumps
that happen to be in the same line of sight.  This is less likely because
channel maps do not show traits of unrelated components.  
If clumps exist in the foreground or background, they are more likely
to show up individually at a peculiar velocity.  
In contrast, the morphology of the envelope at different velocities are
systematically consistent in our observations.
Also, N$_2$H$^+$ does not pick up other velocity components as easily as CO 
because it requires high densities to form. 
Infall motion, not necessarily gravitational infall, is suggested.
%Detail comparison of the velocity structure with more sophisticated
%infall models will be necessary to fully understand this system.

\subsubsection{Rotation} 

While axisymmetric infall cannot explain the velocity difference between 
the east and west extensions of the envelope, the differential velocity 
may come from rotation of the envelope.  
%(talk about the velocity structures predicted by various theoretical
%models...  not gonna test them in this paper though.  Unfortunately,
%the velocity profile cannot be solely determined by the observations.)
In this section, we test some simple rotation curves.  
%using the same toy model of kinematics study.  
%discussed in the previous section.   
%PV diagram and dynamics modeling- a probable rotating collapsing envelope  
%A toy model that illustrates that the picture is consistent or not 
%
%-------Models: 
%(0) central point mass Keplerian (with central mass free-fall velocity?), 
%assuming the central point mass is much 
%larger than the envelope mass. found that the (well-fitted) contours 
%require a very light central mass, implying the invalidity of the assumption. 
%Therefore invalidate this model.  
%(1) pure Keplerian of power-law 2 envelope 
%(2) pure free-fall infall
%(3) rigid-body rotation as the galactic rotation 
%(3) rotation + infall   
%In all of them:
%density profile from the best fit in previous section
%but different inner radius 
%depletion emulated at regions above 10$^6$cm$^{-3}$  
%(therefore the density structure in the inner region does not matter) 
%Beam smoothing reproduced by multiplying a Gaussian of the same FWHM 

\begin{figure}
\includegraphics[angle=270,width=0.6\textwidth]{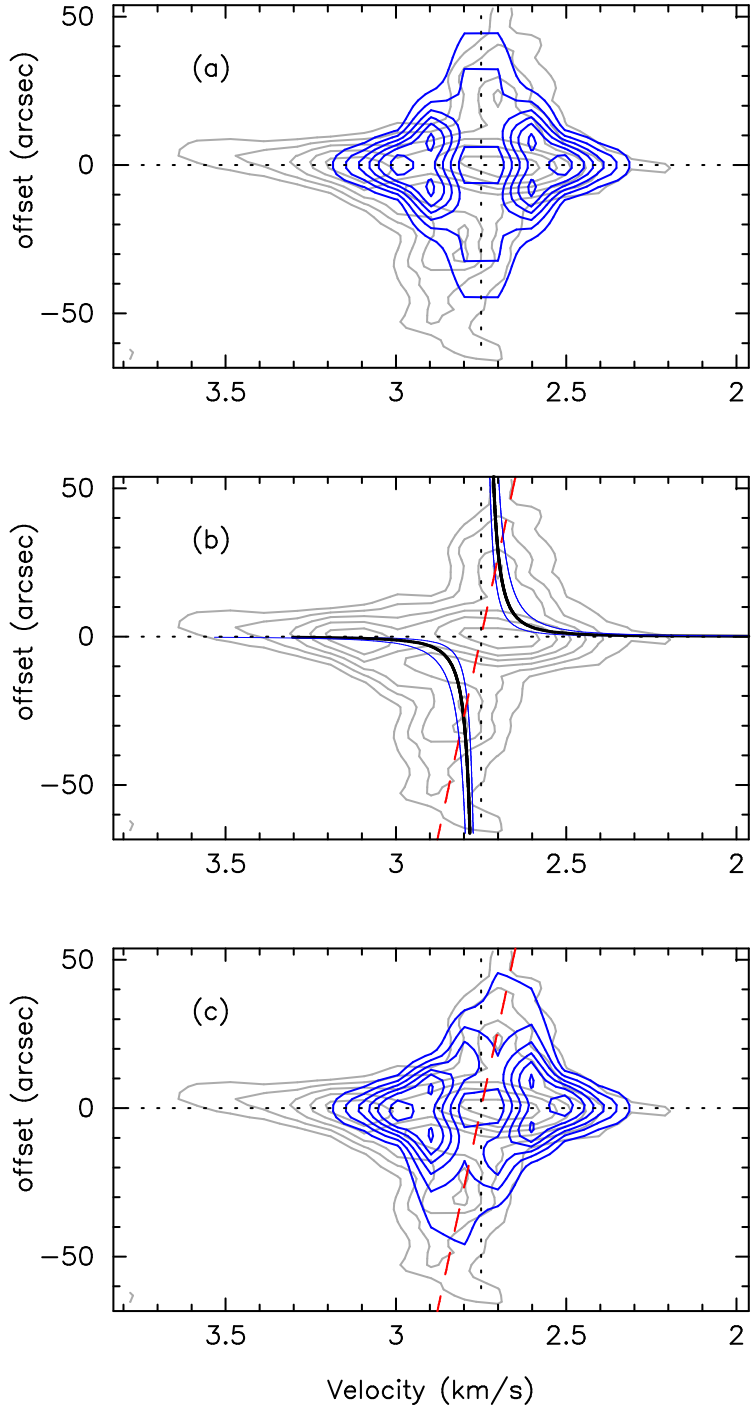} 
\caption{ 
Observed ({\it gray contours}) and model ({\it blue contours}) 
PV diagrams for the flattened envelope of L1157: 
(a) pure infall motion,  
(b) solid curves show the point-mass Keplerian rotation for 
M$_\star$=0.01, 0.02, and 0.04 M$_\odot$, dashed line shows 
solid-body rotation with angular velocity of 1.5 km~s$^{-1}$~pc$^{-1}$ , and 
(c) combined model of infall plus solid-body rotation. 
}
\label{figPVMS}
\end{figure}

First, we test Keplerian rotation. 
If the dynamics of the envelope is dominated by Keplerian rotation  
around a central point mass   
%presumably from the newly-formed protostar, 
much larger than the cumulative envelope mass, the central mass has to be
smaller than 0.1 M$_\odot$ to explain the observed velocity differential
%between the east and west sides 
of the envelope.
In Figure \ref{figPVMS}(b), the pure Keplerian curves for a point mass of 
0.01, 0.02, and 0.04 are shown. 
Apparently, the deduced central mass is around or even smaller than the mass 
of the envelope, which means that the cumulative envelope mass 
cannot be neglected.  
%Keplerian rotation around a central point mass cannot explain the observed velocity structure.   
This is expected because for Class 0 YSOs, mass is mostly distributed in 
the envelope rather than the central protostar \citep[e.g.,][]{Andre1993}.  
Next, we consider complete Keplerian rotation with cumulative envelope mass. 
The velocity is (GM$_R$/R)$^{1/2}$, where M$_R$ is the mass 
contained within the radius R.  
%We use the same envelope model as in dust continuum fitting.  
%Thus a pure power-law of -2 envelope density gives a flat rotation curve. 
An envelope with $\rho \propto r^{-2}$ gives a flat rotation curve. 
The best-fit density obtained in the previous dust continuum fitting gives 
a roughly constant 
rotation velocity 0.5~km~s$^{-1}$, much larger than the observed value. 
%Even with the fiducial Gaussian taper applied, the velocity is still 
%much larger than the observed values. 
%These exercises show that the envelope deviates from Keplerian rotation. 
These exercises show that Keplerian rotation is ruled out 
for the large-scale envelope. 
%(damn Gaussian taper... makes it difficult to have an analytic form of mass distribution.... :(  with the Gaussian taper, the total mass reduced by a factor of around 2.83)

In fact, the system is too young to construct large-scale
Keplerian rotation.  For example, the period of rotation with velocity
0.2 km/s at 1000 AU is $\sim$1.5$\times$10$^5$ yr, larger than the
typical age of Class 0 protostars.  As seen in the PV diagram,
the size scale of rotation is around 10,000AU and requires even longer
dynamical timescale.  Again, the dynamical time estimation suggests the
unsuitability of pure Keplerian rotation for this system.

Another possible scenario is that there exists solid-body rotation in L1157.  
This has been seen in other Class 0 YSOs such as HH212 \citep{Wiseman2001}.  
The solid-body rotation is probably 
inherited from the initial conditions of the large-scale clouds or filaments. 
Angular momentum plays an important role in the protostar evolution  
\citep[e.g.,][]{Bodenheimer1995}. 
In particular, the initial condition of rotation is closely related to 
the core morphology and fragmentation \citep[e.g.,][]{Saigo2008}. 
The extended envelope of L1157 has a velocity gradient of 
around 1.5 km~s$^{-1}$~pc$^{-1}$ %(5.0e-14/s)
assuming solid-body rotation 
(shown by the dashed line in Figure \ref{figPVMS}(b) and (c)).  
%The velocity gradients are also commonly seen in other Class 0 protostars. 
%Chen2007 smaller scale, larger velocity gradients, smaller specific angular momentum. 
The velocity gradient is much smaller than what was found  
in the survey of Class 0 protostars 
\citep[$\sim$ 7 km~s$^{-1}$~pc$^{-1}$,][]{Chen2007}; 
instead, it resembles the typical velocity gradient found in the dense clouds 
\citep[1-2 km~s$^{-1}$~pc$^{-1}$,][]{Goodman1993,Caselli2002}. 
%initial galactic rotation of the clouds? 
Moreover, the rotating N$_2$H$^+$ envelope of L1157 has a size scale 
much larger than the typical size scale of collapsing envelopes.
Both the large size and the slow bulk rotation imply properties more
similar to prestellar cores than collapsing envelopes.
After the protostar has formed in the central densest region, 
the kinematics of the envelope can still be dominated by the remnant rotation 
of the parent dense cloud at large scales, 
while the infall motion takes over at small scales. 
The good alignment of the rotation with other features such as 
flattened geometry and outflow direction also suggest a consistent picture. 

Considering the envelope with radius of 10,000AU and the fitted solid-body 
rotation, the specific angular momentum is around 
4$\times$10$^{-3}$ km~s$^{-1}$~pc.    
We can locate it at a specific angular momentum--rotation radius  plot 
to compare with the other dense cores and protostars  
\citep[e.g.,][]{Ohashi1997b,Belloche2002,Chen2007}.  
%               Figure 6 ; Figure 16 ; Figure 8  
%L1157 has similar properties as IRAM04191 at large scales.?
Although as an infalling envelope at small scales, the large-scale properties 
of the L1157 flattened envelope resemble those of prestellar cores 
more than those of more evolved protostars.  
%Additionally, given the fitted angular velocity at this large size and the
%assumption that this large envelope collapses to form a circumstellar
%disk, the envelope material would fall onto the midplane within a
%centrifugal radius of $r_c = R^4 \Omega^2 / GM \approx$ 500 AU.
Additionally, the fitted angular velocity implies that 
if the large-scale envelope collapses to form a circumstellar disk, 
the envelope material would fall onto the midplane within 
a centrifugal radius of $r_c = R^4 \Omega^2 / GM \approx$ 500 AU.  
%if the envelope material falls onto the midplane to form a circumstellar 
%disk, the falling radius is around 500AU. 
This radius is much larger than the observed T Tauri disks.  
All above implies that the large-scale flattened envelope is probably 
not involved in the dynamical infall activity.   

The interpretation of solid-body rotation is well suited 
from many theoretical points of view. 
For example, magnetic braking can induce solid-body rotation \citep{Basu1994}.  
%regulated by magnetic fields? 
A turbulent core can also yield velocity gradient similar to  
uniform rotation \citep{Burkert2000}.  
%We do not discuss the complexity in this study.  However, 
Here we calculated the $\beta_{rot}$ parameter, defined as  
the ratio of rotational kinetic energy to gravitational energy   
$\beta_{rot} = \frac{\frac{1}{2}I\Omega^2}{qGM^2/R} 
= \frac{p\Omega^2R^3}{2qGM}$, 
at different radii of the flattened envelope.    
(Note that this $\beta_{rot}$ is different from the opacity spectral index 
mentioned in the previous sections.) 
%p/q=2/3 for constant density; p/q=0.22 for $r^{-2}$ sphere 
%(used in Goodman1993 and more )
For L1157, $\beta_{rot}$ is smaller than 2\% throughout the envelope, 
suggesting that the flattened structure is not supported by rotation. 
%(2\% ?)  rotational energy against gravitational energy (Caselli2002?)
%not a rotation-supported disk
%(magnetic support, thermal support, etc..)

The combined best-fit model of simple infall and solid-body rotation
is shown by the black contours in Figure \ref{figPVMS}(c).  
%This is the best PV fit of our toy model.  
Both the double-peaked feature and the large-scale 
velocity gradient can be explained by this model. 
However, the asymmetric features are impossible to model with 
an axisymmetric model in the optically thin case. 
Local clumpiness can be the cause of asymmetry. 
Without further observations with higher spatial and spectral resolutions 
it is hard to justify the detailed properties. 

%Tobin's non-axisymmetric infall?. 

\subsection{Overall Gas-Dust Comparison and Global Picture of L1157} 
%and interpretation...} 

%overall comparison among 8 micron absorption dust, 3mm E-array dust, N$_2$H$^+$ gas

By considering the dust and gas information of Class 0 YSO L1157
altogether, we are able to construct an overall physical picture of
the system.  The dust absorption at 8 $\mu$m from {\it Spitzer} shows a
large-scale extended dense cloud that is flattened perpendicular to the
outflow direction, while the 3 mm dust emission from CARMA D- and E-array
shows a compact spherical structure.  These two observations of dust detect
different components of the envelope.  The extended envelope detected
%by 8 $\mu$m absorption is likely the remnant of the parent cloud, while
by 8 $\mu$m absorption is likely part of the parent cloud, while
the very inner region seen by our dust continuum is the collapsing
envelope.  The flattened geometry may result from physical processes of
core formation.  On the other hand, the N$_2$H$^+$ emission provides a
consistent view both morphologically and kinematically.  The N$_2$H$^+$
feature coincides with the extended dust absorption seen at 8 $\mu$m,
meaning that the same cloud component is observed.  
A slow solid-body rotation at large scales is seen along the major axis 
of elongation, but the flattened structure is not supported by rotation. 
This parent cloud resembles the physical properties of a
prestellar core, while the innermost region is decoupled and undergoes
(gravitational) collapse.

A similar dynamical scenario has also been suggested for another Class 0
object, IRAM 04191+1522 \citep{Belloche2002}.  
%IRAM 30m OTF map. resolution 25\arcsec or 3500AU 
They have obtained a decoupling radius of $\sim$3500AU, which divides the
envelope into a rapidly rotating inner region with free-fall motion and 
a transition region connected to the ambient slowly rotating cloud.  
Their estimated angular velocity at large scales  
($\sim$1.9 km~s$^{-1}$~pc$^{-1}$ at 7000 AU and 
$\lesssim$0.5-1 km~s$^{-1}$~pc$^{-1}$ at 11,000 AU)   
are comparable to what we find for L1157. %the slow rotation in our study.  
More recent interferometric observations revealed faster rotation 
at smaller scales, but resolved out the large-scale structures  
\citep{Belloche2004,Lee2005}. 
%Unfortunately, the single-dish study was restricted by the spatial resolution,
%while more recent interferometric observations usually resolve out the
%large-scale structure \citep[e.g.,][]{Lee2005}.  

%(column density known,assuming a geometry, estimate the volume number density,
%compare with critical density, and depletion density )
%--- This approach may invalidate other candidate scenario.  For example, if the whole extended feature is a flat sheet of similar depth, the volume density of the central part may be too large (?). 
%We construct an edge-on disk-like envelope model and well interpret our 
%data of L1157, although we cannot exclude other complexities.  
We construct an edge-on disk-like envelope model that fits our L1157 data. 
Given the estimated column density and assumptions of envelope geometry,
the volume density averaged along the line of sight can be estimated.
With the assumption of an outer radius of 15,000AU and
%a fixed N$_2$H$^+$ abundance of 2.5$\times$10$^{-10}$, 
a constant N$_2$H$^+$ abundance of 3.0$\times$10$^{-10}$, as used
in our dust modeling, the average gas volume density ranges from
$\sim$2$\times$10$^4$ to $\sim$3$\times$10$^5$cm$^{-3}$.  This volume
density is consistent with the detectable density
of N$_2$H$^+$, that is, it is close to the critical density 10$^5$cm$^{-3}$.   
On the other hand, the average density should not be higher than the 
depletion density ($\sim$10$^6$-10$^7$ cm$^{-3}$); 
this can give a constraint on the thickness of the cloud.  
%On the other hand, the average density should not be higher than the 
%depletion density at around 10$^6$ cm$^{-3}$ \citep{Bergin1997}. 
%This gives a constraint on the thickness of the cloud.  The minimum
%thickness is around 8400AU so that the average volume density for the
%central region would remain smaller than 10$^6$cm$^{-3}$.
%In other words, if the geometry of the N$_2$H$^+$ feature is like a filament, 
%at least the central part of the filament should not be thinner than 8400AU.

\begin{figure}
\includegraphics[angle=270,width=1.0\textwidth]{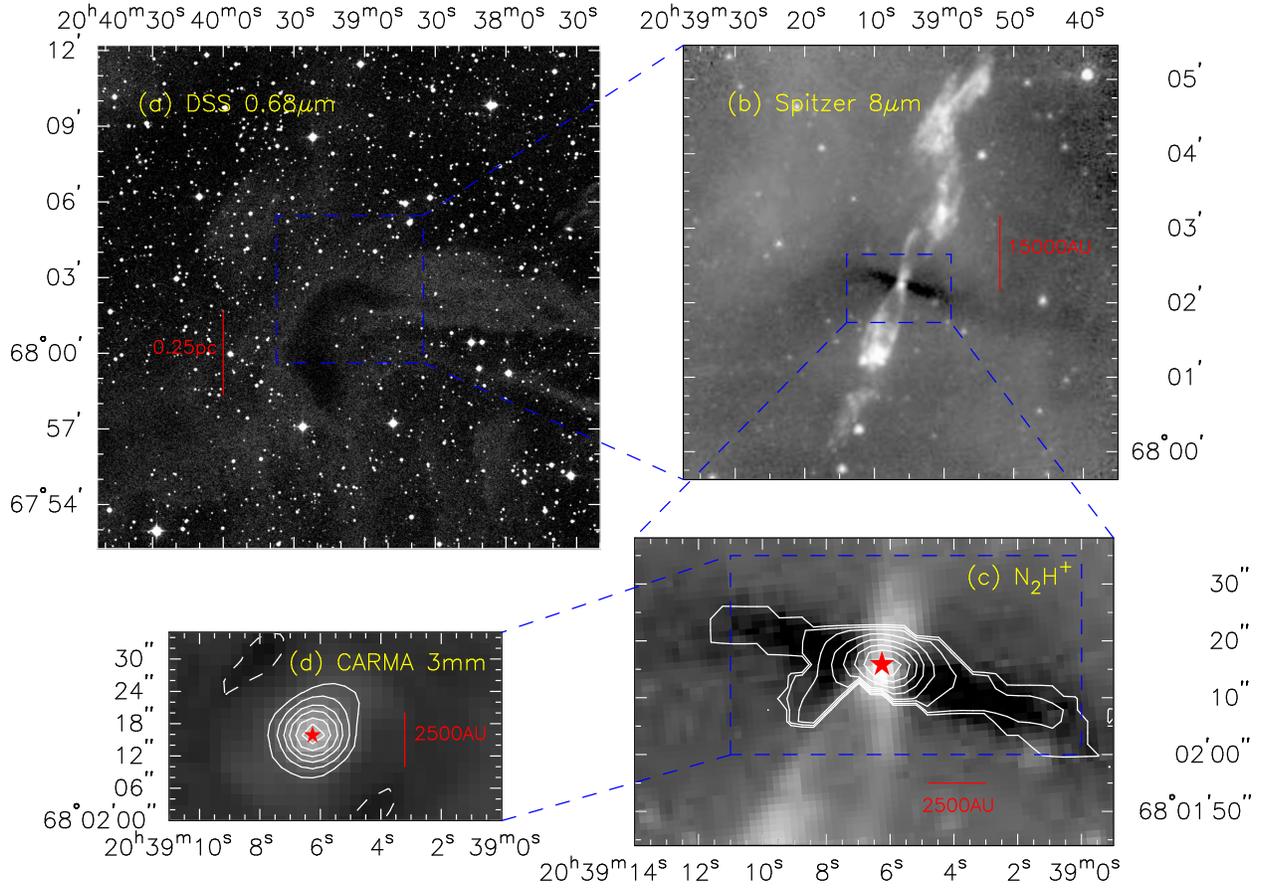} 
\caption{ 
L1157 at different scales seen by various observations:  
(a) DSS optical image, 
(b) {\it Spitzer} IRAC 8 $\mu$m image, 
(c) zoomed-in 8 $\mu$m image overlaid with the N$_2$H$^+$ column density map,  
(d) 3 mm dust continuum observed by CARMA. 
Different geometric characteristics are shown.  
The cloud is irregular at scale of $\sim$0.5 pc, 
flattened perpendicular to the outflows at scale of $\sim$15,000 AU, 
and nearly spherical at size scales smaller than $\sim$5,000 AU.  
}
\label{figDSS}
\end{figure}

%L1157 is an interesting case for early star formation. 
The geometric structures of L1157 at different size scales 
are shown at various observations (Figure \ref{figDSS}).  
At small scales, L1157 shows a nearly spherical morphology, 
as commonly seen in other YSOs  
\citep[e.g.,][]{Looney2000,Shirley2000};
%and is a perfect site for theoretical experiments of collapsing
%protostellar envelope.  So far, 
detailed comparisons between observations and theories 
of collapsing envelopes are usually done assuming spherical symmetry
\citep[e.g.,][]{Looney2003}.
In particular, high-resolution observations done by interferometry 
only show the contributions from the inner envelope while the 
large-scale structures are mostly resolved out. % However, 
Deviation from spherical symmetry can be significant, 
especially at large scales. 
At a younger stage of evolution, large-scale dense cores are commonly
observed to be elongated \citep[e.g.,][]{Myers1991}.
Although the underlying dynamical processes of the flattened cores are
controversial \citep[e.g.,][]{Gammie2003,Tassis2007,Offner2009}, 
non-spherical structure in the initial condition can play an important 
role for their evolution 
\citep[e.g.,][]{Galli1993,Hartmann1996}.   
%Furthermore, many Class 0 YSOs are associated with large-scale 
%non-axisymmetric outer envelopes \citep[e.g.,][]{Tobin2009}. 
For L1157, both the N$_2$H$^+$ feature and 8 $\mu$m absorption reveal
the flattened structure at the size scale of $\sim$0.1 pc, comparable
to a typical prestellar core.
At a larger scale, the morphology becomes even more irregular. 
For example, the Digitized Sky Survey (DSS)  
\footnote{The Digitized Sky Survey was produced at the Space Telescope Science Institute under U.S. Government grant NAG W-2166. The images of these surveys are based on photographic data obtained using the Oschin Schmidt Telescope on Palomar Mountain and the UK Schmidt Telescope. The plates were processed into the present compressed digital form with the permission of these institutions. }
optical image of L1157 shows a large-scale structure of irregular shape.
This is more like the initial condition but not the consequence of
star formation.  
%In Figure \ref{figDSS}, images of L1157 from various
%observations show its geometric structures at different sizescales.
The cloud is filament-like and irregular at very large scale, while
a nearly spherical inner envelope is embedded in the intermediate-size,  
flattened outer envelope extended perpendicular to the outflows. 

%irregular in the largest scale (>10k AU)
%flattened filament in the large scale (10000AU)
%spherical in the small scale (1000sAU)
%presumably protoplanetary disk in the very small scale (<50AU)
%%-------------------------
%L1157 as an example in the role in YSO evolution...  
L1157 is an interesting and relatively simple case for early star formation. 
While many Class 0 YSOs are associated with large-scale 
non-axisymmetric outer envelopes, the geometry of L1157 is 
highly flattened and symmetric \citep{Tobin2009}. 
%With a simpler initial condition, its evolution seems to be less affected by binary formation.  
%as no binary has been found directly so far.  
%but outflow procession Bachiller2001 
L1157 is a perfect site for observational experiments, 
presenting a typical Class 0 YSO with less complexity. 
The properties of L1157 may be generalized to other Class 0 YSOs. 
For example, the Class 0 collapsing envelope with its embedded protostar 
%is still surrounded by their parent core, observed as the outer envelope,  
is surrounded by the outer envelope left by its parent core  
that may or may not gravitationally collapse at a later time. 
In the case of L1157, we are able to detect this outer 
envelope in a flattened structure at both 8 \micron~and N$_2$H$^+$ emission, 
%This stage may be transient since when the inner envelope dominates,  
%temperature effects? the outer envelope dissipates?.. 
revealing a phase that shows both properties of the preceding prestellar core
stage at large scale and properties of the current Class 0 stage at small scale.
In other cases, the outer envelope can become more complicated 
that a non-axisymmetric model will be needed to interpret the observations.  

%A special case that shows the transient moment between two stages 
%at large and small scales. 

%(a schematic view of our interpretation?) 

\section{Summary} 

\begin{enumerate}
\item
We observed the dust continuum and N$_2$H$^+$ gas emission at 3 mm toward 
the Class 0 YSO L1157 IRS with an angular resolution of $\sim$7\arcsec~
using CARMA at D- and E-array.  Spectra of the N$_2$H$^+$ isolated component
$JF_1F$=101-012 were obtained with resolution of $\sim$0.1~km~s$^{-1}$.   
While the 3 mm dust continuum detects the inner envelope, which is 
compact and nearly spherical, the N$_2$H$^+$ emission shows a huge 
flattened structure with a linear size of $\sim$20,000AU, coinciding 
with the disk-like feature found by the 8 $\mu$m absorption in
\citet{Looney2007}.

\item
By fitting the spectra, we estimated the gas column density across the
flattened envelope and compared with the dust column density deduced
from the 8 $\mu$m absorption feature.  
%The results are consistent with previous studies, although some emission is resolved out by interferometry.  
We derived the N$_2$H$^+$ abundance and found
results consistent with what is expected from chemical models.
%Further, we see a slight enhancement of N$_2$H$^+$ abundance towards the
%central condensation, and asymmetry between the east and west extention.
Further, we examined the radial profile of N$_2$H$^+$ abundance along
the major axis of the flattened envelope, 
and showed the asymmetry between the east and west extension.  
%In addition, depletion of N$_2$H$^+$ is expected near the center.

\item
We constructed a simple flattened envelope model that fits the
compact dust continuum; % with statistical significance; 
further, the model-expected gas column density is consistent with the extended
N$_2$H$^+$ emission.  It follows that the deviation from spherical
symmetry can be important at large scales for protostellar envelopes.
However, this model is not motivated by a theoretical model.  
%However, this dust modeling is not intended to justify any detailed
%theoretical structure of the envelope; instead, it provides an example
%of possible scenarios.

\item
We did PV contour modeling and studied the kinematics
of the N$_2$H$^+$ feature.  The spectrum of the central part of the
system shows a double-peaked feature, implying infall.  The large-scale
component can be described by slow solid-body rotation comparable to the
properties of a typical prestellar core.  This large-scale filament
may arguably result from the dynamical processes in the early core
formation, while only the very inner part is actively involved in the
protostar formation.

\end{enumerate}

\acknowledgements{

%The Digitized Sky Survey was produced at the Space Telescope Science Institute under U.S. Government grant NAG W-2166. The images of these surveys are based on photographic data obtained using the Oschin Schmidt Telescope on Palomar Mountain and the UK Schmidt Telescope. The plates were processed into the present compressed digital form with the permission of these institutions.  

%The compressed files of the "Palomar Observatory - Space Telescope Science Institute Digital Sky Survey" of the northern sky, based on scans of the Second Palomar Sky Survey are copyright (c) 1993-2003 by the California Institute of Technology and are distributed herein by agreement. All Rights Reserved. 

H.-F. C. is grateful to C. F. Gammie for insightful discussions and comments.  
We thank the anonymous referee for the valuable comments. 
H.-F. C. and L. W. L. acknowledge support from the Laboratory for Astronomical Imaging at the University of Illinois and the NSF under grant AST-07-09206 and NASA Origins grant NNG06GE41G.
We thank the OVRO/CARMA staff and the CARMA observers for their assistance in obtaining the data. 
Support for CARMA construction was derived from the states of Illinois, California, and Maryland, the Gordon and Betty Moore Foundation, the Eileen and Kenneth Norris Foundation, the Caltech Associates, and the National Science Foundation. Ongoing CARMA development and operations are supported by the National Science Foundation under cooperative agreement AST-0540459, and by the CARMA partner universities. 

}

%\clearpage
\bibliographystyle{apj}
\bibliography{ref}

%--- figures and tables ---

%-----------------------------------------------------------

%-----------------------------------------------------------

%-----------------------------------------------------------
%-----------------------------------------------------------
%-----------------------------------------------------------
%-----------------------------------------------------------
%-----------------------------------------------------------
%-----------------------------------------------------------
%-----------------------------------------------------------
%-----------------------------------------------------------

\begin{deluxetable}{lc}
\tabletypesize{\scriptsize}
%\tablecolumns{2}
\tablewidth{0pc}
\tablecaption{Summary of 3 mm Dust Continuum}
\tablehead{ \colhead{Parameter}       &
            \colhead{Value }          
%\multicolumn{3}{c}{name..}
}
\startdata
Peak flux density (mJy beam$^{-1}$)    \dotfill  & 30.3   \\ 
Noise level $\sigma$ (mJy beam$^{-1}$) \dotfill  & 1.0    \\
Total flux density (mJy)               \dotfill  & 48.3 $\pm$   0.7  \\
%Uncertainty of total flux (mJy)        \dotfill  & 0.7    \\
\sidehead{ Gaussian fit }
%Gaussian fit     \dotfill        &     \\
%  Major axis (arcsec):   \dotfill &     9.364  $\pm$   0.863 \\
%  Minor axis (arcsec):   \dotfill &     8.057  $\pm$   0.392 \\
%  Position angle (degree) \dotfill &    -68.26 $\pm$   18.18 \\
  Major axis (arcsec):   \dotfill &     9.4  $\pm$   0.9   \\
  Minor axis (arcsec):   \dotfill &     8.1    $\pm$   0.4   \\
  Position angle (degree) \dotfill &    -68    $\pm$   18    \\
\sidehead{Deconvolved Size } 
%  Major axis (arcsec)\dotfill&    6.139 \\
%  Minor axis (arcsec)\dotfill&    4.438 \\
%  Position angle (degree)\dotfill&     -53.5
  Major axis (arcsec)     \dotfill &    6.1 $\pm$ 2.3 \\
  Minor axis (arcsec)     \dotfill &    4.4 $\pm$ 0.8 \\
  Position angle (degree) \dotfill &    -54 $\pm$ 21 
\enddata
\tablecomments{
The results of Gaussian fit with the dust continuum done by the MIRIAD task IMFIT. 
}
%\tablenotetext{a}{
%}
%\tablerefs{
%(1) Jennings et al. 1987\nocite{Jennings1987};
%}
\label{tabCont}
\end{deluxetable}

\begin{deluxetable}{cccc}
\tabletypesize{\scriptsize}
\tablecolumns{4}
\tablewidth{0pc}
\tablecaption{Summary of N$_2$H$^+$ Spectrum Fitting }
\tablehead{ \colhead{Parameter}       &
            \colhead{Central Pixel}         &
            \colhead{Mean Value }    &
            \colhead{Mean $\sigma$ }      
%\multicolumn{3}{c}{name..}
}
\startdata
% E-array only 
%$v_{LSR}$ (km~s$^{-1}$) & 2.8333  & 2.8130  & 0.007266 & 2.662 & 2.967  \\
%$\Delta v$(km~s$^{-1}$) & 0.7556  & 0.4588  & 0.01659  & 0.191 & 1.128  \\
%$<\tau>$   & 0.5271  & 0.3394  & 0.01325  & 0.217 & 0.527  \\ 
%Column Density(10$^{12}$ cm$^{-2}$) & 30.061 & 10.65  & 1.775  & 2.138 & 31.77 \\ 
%\cutinhead{ significant digits ...? }
%$v_{LSR}$ (km~s$^{-1}$) & 2.8352 & 2.8190 & 0.0153  \\
%$\Delta v$(km~s$^{-1}$) & 0.7337 & 0.4437 & 0.0349  \\
%$<\tau>$                & 0.4037 & 0.3097 & 0.0255  \\ 
$v_{\mathrm{LSR}}$ (km~s$^{-1}$) & 2.85 & 2.82 & 0.02  \\
$\Delta v$(km~s$^{-1}$) & 0.80 & 0.44 & 0.03  \\
$<\tau>$                & 0.63 & 0.31 & 0.03  \\ 
\sidehead{Derived values :}  
%Column Density(10$^{12}$ cm$^{-2}$) & 22.54 & 9.6616  & 3.2196  \\ 
Column density(10$^{12}$ cm$^{-2}$) & 38.3 & 9.7  & 3.2  \\ 
\enddata
%\tablecomments{
%The uncertainties $\sigma$ are the standard deviations from the MATLAB fitting  
%-- $\Delta v$(FWHM) 
%-- average $\tau$ for the isolated hyperfine component 
%The uncertainty of column density from the assumption of $T_{ex}$ is 
%larger than that from the fitting error propagation. 
%The second row are from temperature (valueAt13K-valueAt7K )/2 
%}
%\tablenotetext{a}{
%}
%\tablerefs{
%}
\label{tabSpec}
\end{deluxetable}
%minimum and maximum are extreme values across the map
%central pixel is (21,13) 
%but the peak column dnsity is at (20,13)

%-----------------------------------------------------------
%-----------------------------------------------------------
%\begin{figure}
%\includegraphics[angle=270,width=1.0\textwidth]{plot.ClnDnVlsr.c.ps} 
%\caption{
%Just wanted to show the color version of Figure \ref{figClnDn}.  
%}
%\end{figure}
%-----------------------------------------------------------
 
%---------------------------------------

\end{document}